\documentclass[journal]{IEEEtran}
\usepackage{cite,url}
\usepackage{rangecite,amsmath,graphicx,epstopdf,amsthm,amssymb,float,color,array,epsfig,subfigure}
\newtheorem{Theorem}{Theorem}
\newtheorem{proposition}{Proposition}
\newtheorem{example}{Example}
\newtheorem{definition}{Definition}
\newtheorem{lemma}{Lemma}
\usepackage{algorithm}
\newcommand{\ud}{\,\mathrm{d}}
\def\upsilonvec{{\mbox{\boldmath $\upsilon$}}}
\newcommand{\avec}{{\bf{a}}}

\newcommand{\bvec}{{\bf{b}}}

\newcommand{\uvec}{{\bf{u}}}
\newcommand{\wvec}{{\bf{w}}}
\newcommand{\xvec}{{\bf{x}}}

\newcommand{\gvec}{{\bf{g}}}

\newcommand{\hvec}{{\bf{h}}}

\newcommand{\etavec}{{\bf{\eta}}}

\newcommand{\zerovec}{{\bf{0}}}

\newcommand{\Lambdamat}{{\bf{\Lambda}}}

\newcommand{\Amat}{{\bf{A}}}
\newcommand{\Bmat}{{\bf{B}}}
\newcommand{\Cmat}{{\bf{C}}}
\newcommand{\Dmat}{{\bf{D}}}

\newcommand{\Fmat}{{\bf{F}}}
\newcommand{\Gmat}{{\bf{G}}}
\newcommand{\Hmat}{{\bf{H}}}
\newcommand{\Jmat}{{\bf{J}}}
\newcommand{\Imat}{{\bf{I}}}

\newcommand{\Pmat}{{\bf{P}}}

\newcommand{\Qmat}{{\bf{Q}}}

\newcommand{\define}{\stackrel{\triangle}{=}}

\def\xsivec{{\mbox{\boldmath $\xi$}}}

\def\etavec{{\mbox{\boldmath $\eta$}}}

\def\thetavec{{\mbox{\boldmath $\theta$}}}

\def\muvec{{\mbox{\boldmath $\mu$}}}

\def\thetavecsmall{{\mbox{\boldmath {\scriptsize $\theta$}}}}

\newcommand{\be}{\begin{equation}}
\newcommand{\ee}{\end{equation}}
\newcommand{\beqna}{\begin{eqnarray}}
\newcommand{\eeqna}{\end{eqnarray}}

\usepackage{algorithm,algorithmic,setspace}
\newcommand{\zp}{^{\text{\tiny{ZP}}}}
\begin{document}
	%	\maketitle
	\title{Cram$\acute{\text{e}}$r-Rao Bound for Estimation After Model Selection and its Application to Sparse Vector Estimation}
	\author{Elad~Meir,~\IEEEmembership{Student Member, IEEE} and 
		Tirza Routtenberg, \IEEEmembership{Senior Member, IEEE}% <-this % stops a space
		\vspace{-0.5cm}
		\thanks{This work is partially supported by  the ISRAEL SCIENCE FOUNDATION (ISF), grant No. 1173/16 and by the BGU Cyber Security Research Center.
		\\
This work has been submitted to the IEEE for possible publication.
Copyright may be transferred without notice, after which this version may
no longer be accessible.}
		\thanks{{\footnotesize{E. Meir and T. Routtenberg are with the School of Electrical and Computer Engineering Ben-Gurion University of the Negev Beer-Sheva 84105, Israel, e-mail: meirela@post.bgu.ac.il, tirzar@bgu.ac.il.}}}% <-this % stops a space
}
	
	\maketitle
	\nopagebreak
	\begin{abstract}
		In many practical parameter estimation problems, such as coefficient estimation of polynomial regression,  {\textcolor{black}{the true model is unknown and thus, a}} model selection step is performed prior to  estimation. The data-based model selection step affects the subsequent estimation. In particular, the oracle Cram$\acute{\text{e}}$r-Rao bound (CRB), which {\textcolor{black}{is based on}} knowledge of the {\textcolor{black}{true}} model, is inappropriate for post-model-selection performance analysis and system design outside the asymptotic region. In this paper, 
	{\textcolor{black}{	we investigate post-model-selection parameter estimation of a  vector with an unknown 
support set, where this support set represents the model.}}
		We analyze the estimation performance of   coherent estimators that force unselected parameters to zero. We use the mean-squared-selected-error (MSSE) criterion and introduce the concept of selective unbiasedness in the sense of Lehmann unbiasedness. We derive a non-Bayesian Cram$\acute{\text{e}}$r-Rao-type bound on the MSSE and on the mean-squared-error (MSE) of any coherent  estimator with a specific selective-bias function in the Lehmann sense. 
		We implement the selective CRB for the special case of sparse vector estimation with an unknown support set. Finally, we demonstrate in  simulations that the proposed selective CRB is an informative lower bound on the performance of the
		 maximum selected likelihood  estimator for a general linear model with the generalized information criterion  and for sparse vector estimation with one step thresholding.
		It is shown that for these cases the selective CRB outperforms the  oracle CRB and
	Sando-Mitra-Stoica CRB
	(SMS-CRB) \cite{Sando_Mitra_Stoica2002}.
	\end{abstract}
	\begin{IEEEkeywords}
		Non-Bayesian selective estimation, 
		selective Cram$\acute{\text{e}}$r-Rao bound, 
		estimation after model selection, 
		coherence estimation,
		sparse vector estimation
	\end{IEEEkeywords}
	%%%%%%%%%%%%%%%%%%%%%%%%%%%%%%%%%%%%%%%%%%%%%%%%%%%%%%%%%%%%%%%%%%%%%%%%%%%%%%%%%%%%%%%%%%%%%%%%%%%%%%%%%%%%%%%	
	\section{Introduction}
	\label{sec:intro}
Estimation after model selection arises in a variety of problems in signal processing, communication, and multivariate data analysis  \cite{Sando_Mitra_Stoica2002,multivariate,7006732,stoica2004model,BnA,WaxKailath,Nadler}.
 In post-model-selection estimation  the common practice is 	to select a model from a pool of candidate models and then,
in the second stage, estimate 
the unknown parameters associated with the selected model.
 For example, in direction-of-arrival (DOA) estimation, first, the number of sources is selected, and then, the DOA of each detected source is estimated \cite{Ottersten1993,Bethel_Bell_2004,Nikolic_Nehorai_Djordjevic2012}. 
The selection in this case is usually based on information theoretic criteria, such as the Akaike’s Information Criterion (AIC) \cite{AIC},
	 the Minimum Description Length (MDL) \cite{MDL}, and the
	 generalized information criterion (GIC) \cite{stoica2004model}.
In  regression models \cite{hurvich1989regression,Francos1},   the significant predictors are identified, and then, 
the corresponding coefficients of the selected model are 
typically estimated by the least squares method.
%%%%%%
Another example is in the context of estimating a sparse unknown parameter vector from noisy measurements.	Sparse estimation has been analyzed intensively in the past	few years, and has already given rise to numerous successful signal processing algorithms (see, e.g. \cite{tropp2006,candes2007,donoho2003optimally}).
	In particular, in greedy  compressive sensing algorithms \cite{CS_Davenport_Eldar,OMP_Mallat93},
the support set of the signals is selected, 
 and then the associated nonzero
values, i.e. the signal coefficients, are estimated. Thus, the problem of non-Bayesian sparse vector recovery can be interpreted as a special case of estimation after model selection.

Post-model-selection estimation procedures
usually set the unselected parameters to zero, and then, estimate the selected parameters by conventional estimators, 
such as  the maximum likelihood (ML) and least squares estimators.
However, the performance of post-model-selection parameter estimation is difficult to evaluate.
The oracle Cram$\acute{\text{e}}$r-Rao Bound (CRB), which {\textcolor{black}{is based on}} perfect knowledge of the {\textcolor{black}{true}} model, is commonly used for performance analysis and system design in these cases (see, e.g. \cite{Ye_Bresler_Moulin2003,sparse_con}).
However, the oracle CRB does not take the prescreening process and the fact 
that the {\textcolor{black}{true}} model is unknown into account and, thus, it is not a valid bound and cannot predict the threshold MSE of nonlinear estimators. A more significant problem is the fact that
the estimation is based on the same dataset utilized in the model selection step.
 The data-driven
selection process
creates ``selection bias" and  produces a model that is itself stochastic, and
this stochastic aspect is not accounted for by classical non-Bayesian estimation theory \cite{ZhaoZhang}.
 For example, it has been shown that ignoring the model selection step  leads to invalid analysis, such as non-covering confidence intervals \cite{potscher1991effects,Leeb_Potscher2005}. 
As a consequence,  classical  MSE lower bounds, such as the oracle CRB, are not valid outside the asymptotic region, nor can they predict the threshold without the help of other measures, such as anomalous error probabilities \cite{athley2005threshold}. 
 Despite the widespread occurrence of estimation after model selection scenarios  in signal processing, the impact of the  selection procedure on the fundamental limits of estimation performance for general parametric models is not well understood.

	%%%%%%%%%%%%%%%%%%%%%%%%%5
\subsection{Summary of results}
	In this paper we investigate the post-model-selection estimation performance 
 for the estimation of a  vector with an unknown 
support set, where this support set represents the model. In this setting,
 the data-based selection rule is given and we analyze the post-model-selection performance for this specific rule.
%The estimated parameters belong to a model that has been selected from a set of candidate models.
We consider the common practice of coherent estimators,
  i.e. estimators that force the unselected parameters to zero. 
		In order to characterize the estimation performance
		of coherent estimators
		we introduce   the mean-squared-selected-error (MSSE) criterion, as a performance measure,
and derive the concept of selective-unbiasedness, by using the  non-Bayesian Lehmann unbiasedness definition \cite{Lehmann}.
Then, we develop  a new post-model-selection Cram$\acute{\text{e}}$r-Rao-type lower bound, named selective CRB,
on the MSSE of any coherent 
estimator with a given   selective  bias in the Lehmann sense.
As a special case, we implement the proposed selective CRB for  the problem of sparse vector recovery  from a small number of noisy measurements.
	The selective CRB is examined in simulations for a linear regression problem and for sparse vector recovery
	and is shown in both to be
	 an informative bound also outside the asymptotic region, while the oracle CRB is not, and to be tighter than the Sando-Mitra-Stoica CRB  
	(SMS-CRB) in \cite{Sando_Mitra_Stoica2002}. 

\subsection{Related works}
The majority of work on selective inference in mathematical statistics literature is concerned with constructing confidence intervals \cite{ZhaoZhang,potscher1991effects,Leeb_Potscher2005,Lee_Taylor_2014,EfronX,Benjamini2005false,Kabaila_Leeb_2006,TibsAs,Rosenblatt2014401,Tibshirani_Taylor_Lockhart_Tibshirani2016}, testing after model selection 
\cite{2014arXiv1410.2597F,Heller_Meir_Chatterjee2017},
and post-selection ML estimation \cite{meir2017tractable,Heller_Meir_Chatterjee2017}. These works were usually {\textcolor{black}{developed for}} specific models, such as linear models, and specific estimators, such as M-estimators \cite{potscher1991effects}  or the Lasso method \cite{meir2017tractable}. 
Here, we provide a general {\em{non-Bayesian}} estimation framework
for any parametric model and  any estimator.

	In the context of signal processing, 
	 in
\cite{hero2}  Bayesian estimation after the detection of an unknown data region of interest has been investigated.
However, in this case
the useful {\em{data}} is selected and not the {\em{model}}.
 Bayesian post-model selection has been investigated in  \cite{Harel_Routtenberg_Bayesian}.
In the non-Bayesian framework, a novel CRB on the conditional mean-squared-error (MSE)  is developed in  \cite{Chaumette2005,energy} for the problem of post-detection estimation.
 In \cite{Weiss_Routtenberg_Messer}, a new
performance evaluation measure for the post-detection estimation  of an intensity parameter that incorporates the
estimation- and detection-related errors is proposed.
The effects of random compression on  the CRB have
been studied  in \cite{Pakrooh_Scharf_Howard_2015}.
%%%
MSE approximations by the method of interval errors  \cite{athley2005threshold,Richmond2006,richmond2012aspects}
	indicate that  the threshold phenomenon of the MSE is highly related to the probabilities of different errors \cite{Merhav_2011}.
		In \cite{Routtenberg_Tong_ICASSP14,RoutTong2016,Harel_Routtenberg_2019},
		we developed the CRB and estimation methods for models with nuisance unknown  parameters, whose ``parameters of interest"  are selected  based on the data,
		i.e.  estimation after \textit{parameter} selection, in which 
	the true model is perfectly known.
	In contrast,
 in the case presented here, 
the true measurement model is  {\em{unknown}}
 and is selected from a finite collection of competing models.
 %Thus, the bound from 
%\cite{Routtenberg_Tong_ICASSP14,RoutTong2016,Harel_Routtenberg_2019}, 
%as well as post-parameter-selection estimation methods, are irrelevant for estimation after {\em{model}} selection.
	%In the considered scenario, however,  we know the full finite set of candidate models that can be assumed. 
	%Thus, in the proposed approach the estimation errors are from specific categories and can be averaged along these models. 
{\textcolor{black}{The	CRB for general estimation under a misspecified  model has been developed in \cite{richmondHorowitz,vuong,pajovic,Fortunati_Gini_Greco2016}. 
However, the misspecified CRB is  a lower bound on the MSE of the ``pseudo-true parameter vector”, which minimizes the Kullback–Leibler divergence between the true  and assumed distributions, and not on the true parameter, as in this paper. }}

In the pioneering work of Sando, Mitra, and Stoica in \cite{Sando_Mitra_Stoica2002},  a novel CRB-type bound is presented for estimation after  model order selection, named here as SMS-CRB. The SMS-CRB is based on some restrictive assumptions on the selection rule and on averaging the   Fisher information matrices (FIMs) over the different models. As a result, it is not a tight bound, as presented in the simulations herein. 
	For the special case of sparse vector estimation, 
the associated constrained CRB  \cite{gorman1990,stoica1998} is reduced to the oracle CRB  \cite{sparse_con,babadi2009},
which {\textcolor{black}{is based on  perfect  knowledge of the true}} support set and is non-informative outside the asymptotic region.

%%%%%%%%%%%%%%%
\subsection{Organization and notations}
	The remainder of the paper is organized as follows: Section \ref{probFor} presents the mathematical model for the problem of estimation	after model selection. In Section \ref{sCRB} the proposed selective CRB is derived, with its different versions.
	In Section \ref{sparse}, we implement the selective CRB for  sparse vector estimation. The performance of the proposed bound  is evaluated in simulations in Section \ref{simulations} and our conclusions can be found in Section \ref{conclusion}.
	%%%%%%%%%%%%%%%%%%%%%%%%%%
	
	In the rest of this paper, we denote vectors by boldface lowercase letters and matrices by boldface uppercase letters. 
	The operators  $(\cdot)^T$, $(\cdot)^{-1}$,  and ${\text{Tr}}(\cdot)$ denote the transpose, inverse,  and trace operators, respectively.
For a matrix $\Amat\in{\mathbb{R}}^{M\times K}$ with a full column rank,
		$\Pmat_\Amat^\bot=\Imat_M-\Amat(\Amat^T\Amat)^{-1}\Amat^T$,
		%%%%%is the orthogonal projection matrix onto  the null space of $\Amat^T$, 
		where $\Imat_M$ is the  identity matrix of order $M$.
	The $m$th element of the vector $\avec$,
	the $(m,q)$th element of the matrix $\Amat$,
	and the $(m_1:m_2\times q_1:q_2)$ submatrix of $\Amat$ 
	are denoted by $a_m$, $\Amat_{m,q}$, and 
	$\Amat_{m_1:m_2,q_1:q_2}$, respectively. 
	For two symmetric  matrices  of the same size
	$\Amat$ and $\Bmat$, $\Amat\succeq\Bmat$ means that $\Amat-\Bmat$ is a positive-semidefinite matrix. 
	 The gradient of a vector function, $\gvec$, of $\thetavec$, $\nabla_{\thetavecsmall}\gvec$, is a matrix in $\mathbb{R}^{K\times M}$, with the $(k,m)$th element equal to $\frac{\partial g_k}{\partial \theta_m}$, where $\gvec=\left[g_1,\ldots,g_K\right]^T$ and $\thetavec=\left[\theta_1,\ldots,\theta_M\right]^T$. For any index set, $\Lambda \subset \{1,\ldots,M\}$,
	$\thetavec_{\Lambda}$ is the $|\Lambda|$-dimensional subvector of $\thetavec$ containing the elements indexed by $\Lambda$, where  $|\Lambda|$ and $\Lambda^c \define \{1,\ldots,M\}\backslash\Lambda$ denote the set's cardinality and complement set, respectively. 
	In addition,  ${\cal{P}}(\{1,\ldots,M\})$ denotes the power set of the set $\{1,\ldots,M\} $.
	The notation $\Amat_\Lambda$ stands for a submatrix of $\Amat$ consisting of the columns indexed by $\Lambda$, and
	 ${\mathbf{1}}_{\mathcal{A}}$ denotes the indicator function of an event $\mathcal{A}$.

	%Finally, $\evec_m$ is a vector (with appropriate dimension) of zeros, except for the $m$th element, which is $1$.

	%%%%%%%%%%%%%%%%%%%%%%%%%%%%%%%%%%%%%%%%%%%%%%%%%%%%%%%%%%%%%%%%%%%%%%%%%%%%%%%%%%%%%%%%%%%%%%%%%%%%%%%%%%%%%%%	
	\section{Estimation after model selection}
\label{probFor}
Let $\thetavec=[\theta_1,\ldots,\theta_M]^T\in{\mathbb{R}}^M$ be an unknown deterministic parameter vector. 
{\textcolor{black}{In estimation after model selection, 
the random observation vector, $\xvec \in \Omega_\xvec$, is truly}} distributed according to the probability density function (pdf) $f(\xvec;\thetavec_\Lambda)$,
	in which 
	 $\Omega_\xvec$ is the observation space and
	$\Lambda \in {\cal{P}} (\{1,\ldots,M\})$ is an unknown
		  set of indices.
	This set	  
represents the unknown {\em{model}} and 
 is called the support set in the following. 	
 {\textcolor{black}{In addition, it is known {\em{a-priori}} that the true support set, $\Lambda$, is one of the candidate support sets, 
 $\Lambda_k$, $k=1,\ldots,K$, i.e. $\Lambda\in  \{\Lambda_k\}_{k=1}^K$,}}
 where each support set satisfies  $\Lambda_k\in {\cal{P}} (\{1,\ldots,M\})$
 and represents a different candidate model.
 	In particular, 
	$1\leq|\Lambda_k|\leq M$, $\forall k=1,\ldots,K$,
	{\textcolor{black}{and, thus, the true support set of the vector satisfies $1\leq |\Lambda|\leq M$. In this setting, the full parameter vector, $\thetavec\in{\mathbb{R}}^M$, represents the set of all possible parameters that could parameterize (but may not parameterize) the true distribution.
	{\textcolor{black}{Therefore,
	 although the true pdf of $\xvec$ is unknown, it is  known that the true parameter vector that parameterizes this pdf, $\thetavec_\Lambda$, will be chosen as a subset of the  vector $\thetavec$.
	}}
	As a result, we can say that }} 
 the true pdf, $f(\xvec;\thetavec_{\Lambda})$,
{\textcolor{black}{belongs to a given}} set of pdfs,
	$\{f(\xvec;\thetavec_{\Lambda_k})\}$, $k=1,\ldots, K$. Each pdf in this set is parameterized by its own unknown parameter vector, $\thetavec_{\Lambda_k}$.
	In the following, the notation ${\rm{E}}_{\thetavecsmall_\Lambda} [\cdot]$ and $ {\rm{E}}_{\thetavecsmall_\Lambda} [\cdot|\mathcal{A}]$ represent the  expected value and the conditional expected value, computed  by using the true pdf, $f(\xvec;\thetavec_{\Lambda})$.  

\begin{figure}[hbt]
		\begin{center}	
		\includegraphics[scale=0.16]{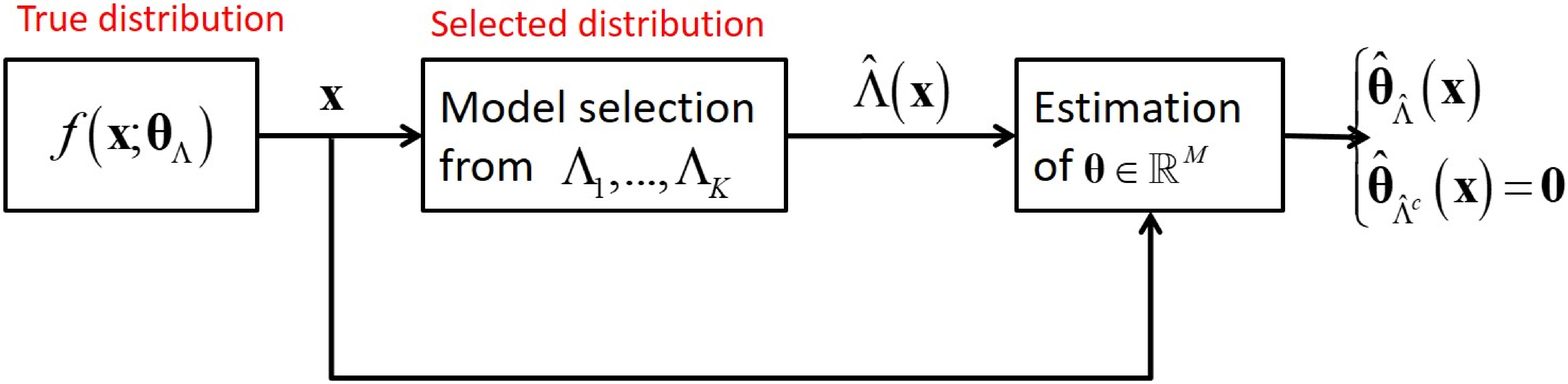}
		\caption{Estimation after model selection:
		{\textcolor{black}{The true pdf that generates
		the measurement vector, $\xvec$, is  $f(\xvec;\thetavec_\Lambda)$, where {\textcolor{black}{ $\Lambda\in \{\Lambda_k\}_{k=1}^K$ is the true support set, such that
		$\Lambda_k\in {\cal{P}} (\{1,\ldots,M\})$, $k=1,\ldots,K$.}}
		In the first processing stage, a model (support set) is selected according to a predetermined selection rule, which results in an estimated support set, $\hat{\Lambda}\in \{\Lambda_k\}_{k=1}^K$.}} In the second stage, the {\textcolor{black}{full}} unknown parameter vector, $\thetavec\in{\mathbb{R}}^M$, is estimated based on  $\xvec$ and on  $\hat{\Lambda}$, where coherency implies that $\hat{\thetavec}_{\hat{\Lambda}^c}=\zerovec$. }
		\label{systemGraph}	
		\end{center}
	\end{figure}

Let $\hat{\thetavec}=[\hat{\theta}_1,\ldots,\hat{\theta}_M]^T$ be an estimator of $\thetavec$,
based on a random observation vector, $\xvec\in\Omega_\xvec$,
i.e.  $\hat{\thetavec}:\Omega_\xvec\rightarrow{\mathbb{R}}^M$, with a bounded second moment. 
	Since 
{\textcolor{black}{the true support set}}, $\Lambda$, is unknown,
	a model selection approach is conducted before the estimation. We take this model selection for granted and analyze the consequent estimation. Thus, estimation after model selection consists of two stages: first,
 a certain model is selected according to a predetermined data-driven selection rule, which results in an estimated support set, $\hat{\Lambda}:\Omega_\xvec \rightarrow 
 	{\textcolor{black}{ \{\Lambda_k\}_{k=1}^K}}$.
 Then, in the second stage, 	the vector of  parameters
 that belong to the selected support set, $\thetavec_{\hat{\Lambda}}$,  is estimated based on the same observation vector, $\xvec$.  We assume here  the  usual practice in post-model-selection estimation, which  is to force the  unselected parameters to zero. The following is a formal definition for this commonly-used practice, named here ``coherency", which is defined with respect to (w.r.t.) the selection rule.
	\begin{definition}
		\label{cohdef}
		An estimator, $\hat{\thetavec}\in{\mathbb{R}}^M$, is said to be a coherent estimator of $\thetavec$ w.r.t. the selection rule, $\hat{\Lambda}$, if
		\be
		\label{coherency_gen} 
		\hat{\thetavec}_{\hat{\Lambda}^c}=\zerovec.
		\ee
	\end{definition}
Finally, the approach of
	estimation after model
selection is presented schematically in Fig. \ref{systemGraph}.
	\hspace{-0.25cm}
	\vspace{-0.5cm}
	
	The probability of selecting the $k$th model is denoted as 
	\be
	\label{pi_k}
\pi_{k}(\thetavec_\Lambda)\define
 \Pr({\hat{\Lambda}=\Lambda}_k;\thetavec_\Lambda),~k=1,\ldots,K,
\ee
where this probability is computed w.r.t.
 the {\em{true}} pdf, $f(\xvec;\thetavec_\Lambda)$. We assume that the deterministic sets
	\be
	\label{Akk}
	{\mathcal{A}}_{k}\define \{\xvec: \xvec\in\Omega_\xvec,\hat{\Lambda}(\xvec)=\Lambda_k\}, ~k=1,\ldots,K,
	\ee
	generate a partition of $\Omega_\xvec$. 
That is, 	$\{{\mathcal{A}}_{k}\}_{k=1}^K$ are  mutually disjoint subsets of $\Omega_\xvec$,  whose union is $\Omega_\xvec$.
	By using Bayes rule, it can be verified that
	\be
	\label{bayes}
		f(\xvec |{\hat{\Lambda}=\Lambda}_k; \thetavec_\Lambda)=\frac{f(\xvec;\thetavec_\Lambda)}{\pi_k(\thetavec_{\Lambda})}, ~\forall \xvec \in {\mathcal{A}}_k,	
	\ee
	% for any
	$\forall {\mathcal{A}}_k$ such that $\pi_k(\thetavec_{\Lambda})\neq 0$.  
	%For ${\mathcal{A}}_k$ with zero probability we arbitrarily set	$f(\xvec |{\hat{\Lambda}=\Lambda}_k; \thetavec_\Lambda)=0$.
We also define the null parameters that have been {\em{wrongly}} selected by the  selection rule.
%%%as follows.
		\begin{definition}
		\label{null_def}
	{\textcolor{black}{	A parameter is said to be a null parameter if it does not appear in the true model.  Thus, the set of all null parameters is given by $\thetavec_{\Lambda^c}$.}}
	\end{definition}

	The support sets of the unknown parameter vectors, $\{\Lambda_k\}_{k=1}^K$, {\textcolor{black}{and the true support set, $\Lambda$,}} differ in size. In order to compare between the estimation errors in different models, we introduce the zero-padded vectors, where the zero-padding  is  to the length of the full parameter vector, $M$. 	
	\begin{definition}
	\label{ZPdef}
		For an arbitrary vector, $\avec\in{\mathbb{R}}^M$, and any candidate support set, ${\Lambda}_k$, $k=1,\ldots,K$, the vector 
		$\avec_{{\Lambda}_k}\zp$, is a zero-padded, $M$-length vector, whose non-padded elements correspond to the elements of $\avec_{{\Lambda}_k}$.
	\end{definition}

In this paper, we are interested in analyzing the performance of coherent estimators, as defined in Definition \ref{cohdef}.	
	Thus, only estimation errors that belong to  the estimated (non-padded) parameters are relevant in the resultant zero-padded vector.
This definition is based on the considered scheme, in which
	 the selection rule is predetermined,
	 and our goal is to analyze the post-model-selection estimation approach.
	 %and  is assumed to have been designed   in order to choose the parameters that truly play a role in the data generating distribution.
{\textcolor{black}{	 
In particular, it can be verified that for any coherent estimator, as defined in Definition \ref{cohdef}, $\hat{\thetavec}_{\hat{\Lambda}}\zp=\hat{\thetavec}$.
The following example demonstrates  our notations.
	\begin{example}
Consider a case where there are $M=5$
candidate parameters,
 i.e. $\thetavec\in{\mathbb{R}}^5$, and the true support set and true parameter vector are
$\Lambda=\{1,2,3\}$ and $\thetavec_{\Lambda}=[\theta_1,\theta_2,\theta_3]^T$, respectively. Let us assume that
 for a specific realization (i.e. specific observation vector, $\xvec$), the selection rule results in the estimated support set  $\hat{\Lambda}=\{1,3,5\}$.
Then, according to Definition \ref{cohdef}, any coherent  estimator has the form
$\hat{\thetavec}=[\hat{\theta}_1,0,\hat{\theta}_3,0,\hat{\theta}_5]^T$.
According to Definition \ref{ZPdef}, the zero-padded estimation error vector for this case is
$\hat{\thetavec}_{\hat{\Lambda}}\zp
-\thetavec_{\hat{\Lambda}}\zp=
[
\hat{\theta}_1-\theta_1,0,\hat{\theta}_{3}-\theta_{3},0,\hat{\theta}_5]^T$. Thus, for this observation vector, estimation errors include the influence of  estimating the true parameters that have also been selected, $\theta_1$ and $\theta_3$, as well as the influence of the null parameter, $\theta_5$.
	\end{example}}}

	%%%%%%%%%%%%%
We use the following selected-square-error (SSE) matrix cost function:
	\beqna
	\label{aSSE}
		\Cmat(\hat{\thetavec},\hat{\Lambda},\thetavec)\define\left(\hat{\thetavec}_{\hat{\Lambda}}\zp-\thetavec_{\hat{\Lambda}}\zp\right)\left(\hat{\thetavec}_{\hat{\Lambda}}\zp-\thetavec_{\hat{\Lambda}}\zp\right)^T.
	\eeqna
	The corresponding mean SSE (MSSE) is given by
	\beqna
	\label{MSSE}
		{\rm{E}}_{\thetavecsmall_\Lambda}\left[\Cmat(\hat{\thetavec},\hat{\Lambda},\thetavec)\right]
		={\rm{E}}_{\thetavecsmall_\Lambda}\left[\left(\hat{\thetavec}_{\hat{\Lambda}}\zp-\thetavec_{\hat{\Lambda}}\zp\right)\left(\hat{\thetavec}_{\hat{\Lambda}}\zp-\thetavec_{\hat{\Lambda}}\zp\right)^T\right]\nonumber\hspace{0.5cm}
\\
		=\sum_{k=1}^K\pi_{k}(\thetavec_{\Lambda})
		{\rm{E}}_{\thetavecsmall_\Lambda}\left[(\hat{\thetavec}_{\Lambda_k}\zp-\thetavec_{\Lambda_k}\zp)
(\hat{\thetavec}_{\Lambda_k}\zp-\thetavec_{\Lambda_k}\zp)^T|{\hat{\Lambda}=\Lambda}_k\right],
	\eeqna
	where the last equality is obtained by using (\ref{bayes}) and the law of total expectation.
	
	The marginal MSSE on a specific parameter, $\theta_m$, is given by the $m$th diagonal element of the MSSE, that is,
		\beqna
	\label{MSSE_marg}
		{\rm{E}}_{\thetavecsmall_\Lambda}\left[[\Cmat(\hat{\thetavec},\hat{\Lambda},
		\thetavec)]_{m,m}\right]
		\hspace{4cm}\nonumber\\=
		\sum_{k\in \kappa_m}\pi_{k}(\thetavec_{\Lambda}) 
		{\rm{E}}_{\thetavecsmall_\Lambda}\left[(\hat{\theta}_{m}-\theta_{m})^2|{\hat{\Lambda}=\Lambda}_k\right],
	\eeqna
	$ \forall m=1,\ldots,M$, where
	\be
	\label{kappa1}
	\kappa_m\define \{k|k=1,\ldots,K,~m\in\Lambda_k\}
	\ee
	is the set of all the models in which the parameter $m$ is a part of the support set.
	Similarly,
	$\kappa_m^c\define \{k|k=1,\ldots,K,~m\notin\Lambda_k\}$
	is the set of indices of all models for which the parameter $\theta_m$ is not in the associated support set $\Lambda_k$, and, therefore, its value is   zero.
	It can be seen that
	\be
	\label{pm_def}
	p_m(\thetavec_{\Lambda})\define\sum\nolimits_{k\in \kappa_m}\pi_{k}(\thetavec_{\Lambda}) 
	\ee
	is the probability that the parameter 
	$m$ has been selected by the considered selection rule. Thus,
	by using the law of total expectation, (\ref{MSSE_marg}) can be written as
	\beqna
		\label{MSSE_marg2}
	{\rm{E}}_{\thetavecsmall_\Lambda}[[\Cmat(\hat{\thetavec},\hat{\Lambda},\thetavec)]_{m,m}]
		=p_m(\thetavec_{\Lambda}) 
		{\rm{E}}_{\thetavecsmall_\Lambda}[(\hat{\theta}_{m}-\theta_{m})^2|m \in \hat{\Lambda}].
	\eeqna
		It should be noted that    $\theta_{m}=0$ in \eqref{MSSE_marg} and \eqref{MSSE_marg2} for any $m\notin \Lambda$.

	It can be seen that the SSE cost function from (\ref{aSSE}) only takes into account the estimation errors of the elements  that are  not forced to zero by the selection stage. The rationale behind this cost function is that the estimation errors of the  unselected parameters  are only determined by the selection rule, and cannot be reduced by any coherent estimator. Thus, designing and analyzing post-model-selection estimators  can be done  only w.r.t. the estimation errors of the selected parameters  that can be controlled. 
	 In our preliminary work in \cite{meirRoutSSP}, we consider a limited SSE cost where
the estimation errors taken over 	the intersection of the true and estimated support. In this paper, 
	the MSSE  integrates  the estimation errors  of all selected parameters, that is,  {\textcolor{black}{both the true parameters on the support set, $\Lambda$, and the second moment of the null parameters, as defined in Definition \ref{null_def}. The rationale 
	behind also taking into account the
 influence of the null parameters is since, in practical applications, the MSE of these selected null parameters significantly affects the estimation results, especially in the sense of the bias function. In addition, the estimators of the null parameters affect the estimators of the true parameters. Thus, the correlations between the two types of errors via the matrix cost function (i.e. the off-diagonal terms of the SSE in \eqref{aSSE}) should also be taken into consideration. Finally, it can be seen that ignoring the contribution of the null parameters' estimation errors to the risk function may lead to estimators of the null parameters that can be arbitrarily large, even though these parameters do not belong to the model. Thus, we want to control their MSE via the associated second moments. }}

	%%%%%%%%%%%%%%%%%%%%%%%%%%%%%%%%%%%%%%%%%%%%%%%%%%%%%%%%%%%%%%%%%%%%%%%%%%%%%%%%%%%%%%%%%%%%%%%%%%%%%%%%%%%%%%%
\section{Selective CRB}
\label{sCRB}
	In this section, a  CRB-type lower bound for estimation after model selection is derived. The proposed  bound is a lower bound on the MSSE of any coherent  estimator,
	as a function of its selective bias function,
	where selective unbiasedness is defined in Section \ref{sUnbias} by using
	%the concept of 
	Lehmann unbiasedness.
	Section \ref{relation} shows the relation between the MSE and the MSSE for coherent estimators as a function of the selective bias.
	The  selective CRB is presented in Section \ref{sCRBsec}, followed by  remarks and  special cases in Section \ref{specialCases}. An preliminary derivation of the scalar selective CRB  appears in \cite{meirRoutSSP}.
%Since in this work, we take the selection rule for granted and analyze the post-model-selection performance, the proposed selective CRB and selective unbiasedness are functions of the specific model selection rule.  
	
	%%%%%%%%
	\subsection{Selective unbiasedness}
	\label{sUnbias}
	In order to exclude trivial estimators, the mean-unbiasedness constraint is commonly used in non-Bayesian parameter estimation \cite{Kay_estimation}. However, this constraint is inappropriate for estimation after model selection, since we are interested only in errors of the selected parameters and since the data-based model selection step induces bias \cite{ZhaoZhang}. Lehmann \cite{Lehmann} proposed a generalization of the unbiasedness concept based on the considered cost function. In our previous work (p. 13 in \cite{RoutPhd}) we extended the scalar Lehmann unbiasedness definition to the general case of a matrix cost function, as follows.
	\begin{definition}
	\label{Ldef}
		The estimator, $\hat{\thetavec}$, is said to be a uniformly unbiased estimator of $\thetavec$ in the Lehmann sense w.r.t. the positive semidefinite matrix cost function, $\Cmat(\hat{\thetavec},\thetavec)$, if
		\be
			\label{Lehmann_vector}
			{\rm{E}}_{\thetavecsmall}[\Cmat(\hat{\thetavec},\etavec)] \succeq {\rm{E}}_{\thetavecsmall}[\Cmat(\hat{\thetavec},\thetavec)],~\forall \etavec,\thetavec\in \Omega_\thetavecsmall,
		\ee
		where $\Omega_\thetavecsmall$ is the parameter space.  
	\end{definition}
	Lehmann unbiasedness conditions for various cost functions can be found in 
	\cite{Routtenberg_Tong_ICASSP14,RoutTong2016,PCRB_J,Routtenberg_cyclic,Eyal_constraint}.
	%%%
	The following proposition defines a sufficient condition for the selective unbiasedness property of estimators w.r.t. the SSE matrix cost function and the selection rule.
	To this end, we define 
	the selective bias
		of the  estimator, $\hat{\thetavec}$,
		under the selected $k$th model and the selection rule, $\hat{\Lambda}$, as follows:
	\be
		\label{11_bias}
		\bvec_k(\thetavec,\Lambda)\define
		{\rm{E}}_{\thetavecsmall_\Lambda}[\hat{\thetavec}_{\Lambda_k}\zp-\thetavec_{\Lambda_k}\zp|{\hat{\Lambda}=\Lambda}_k],~ k=1,\ldots,K.
		\ee
	\begin{proposition}
		\label{propLehm}
	 If 
\be
		\label{11_bias_zero}
		\bvec_k(\thetavec,\Lambda)=\zerovec,~ k=1,\ldots,K,
		\ee 
		such that $\pi_k(\thetavec_{\Lambda})\neq 0$,
then, the estimator $\hat{\thetavec}$ is an unbiased estimator of  $\thetavec$ in the Lehmann sense  w.r.t. the SSE matrix defined in (\ref{aSSE})
		and the selection rule $\hat{\Lambda}$.
	\end{proposition}
	\begin{IEEEproof} The proof appears in Appendix \ref{unbiasApp}.
	\end{IEEEproof}
	It should be noted that 
	%the selective unbiasedness restricts only the values that belong to the estimated support set to  be unbiased.
	%	Thus, 
		the selective unbiasedness is defined as a function of the specific model selection rule.
		In the following, an estimator, $\hat{\thetavec}$, is said to be a selective unbiased estimator for the problem of estimating $\thetavec$ with the support set, $\Lambda$, and a given model selection rule, $\hat{\Lambda}$, if the
	sufficient condition in \eqref{11_bias_zero} is satisfied.
%According to \eqref{11_bias} and to
%Proposition \ref{propLehm},
%the condition in \eqref{11_bias_zero} is a sufficient condition for the Lehmann unbiasedness.
However, it should be noted that 
in practice, coherent post-model-selection estimators tend to be biased,
 similarly to in  various cases of nonlinear estimation problems \cite{SomekhBaruch_Leshem2017}.
		
		The condition in \eqref{11_bias_zero} implies the requirement that
all the scalar estimators of the elements of $\thetavec$
satisfy
	\be
	\label{11_for_m2}
		{\rm{E}}_{\thetavecsmall_\Lambda}[\hat{\thetavec}_{m}|\hat{\Lambda}=\Lambda_k]=\left\{\begin{array}{lr}
		\theta_{m} &k\in 	\kappa_m\\
	0	&k\notin 	\kappa_m
	\end{array}\right.
	,~\forall m =1,\ldots,M,
	\ee
	$\forall \thetavec\in{\mathbb{R}}^{M}$ and $\Lambda\subseteq\{1,\ldots,M\}$.
		It should be noted that   $\theta_{m}=0$ in  \eqref{11_for_m2}  for the null parameters.
	In addition, by  multiplying  (\ref{11_bias_zero}) by $\pi_{k}(\thetavec_{\Lambda})$ and  summing over the candidate models,  $k=1,\ldots,K$, we obtain that the  condition  in \eqref{11_bias_zero} implies, in particular, that $\forall \thetavec\in{\mathbb{R}}^{M}$, $\Lambda\subseteq\{1,\ldots,M\}$, 
\be
\label{s_unbiased}
\sum\nolimits_{k=1}^K\pi_{k}(\thetavec_\Lambda)\bvec_k(\thetavec,\Lambda)={\rm{E}}_{\thetavecsmall_\Lambda}[\hat{\thetavec}_{\hat{\Lambda}}\zp-\thetavec_{\hat{\Lambda}}\zp]=\zerovec.
\ee

	%%%%%%%%%%%%%%
	\subsection{Relation between MSE and MSSE}
	\label{relation}
	In this subsection we develop the relation between the MSE and MSSE measures for coherent estimators.	This relation is used in Subsection \ref{sCRBsec} for obtaining a lower bound on the MSE of
	coherent  estimators directly from the selective CRB
	on the MSSE.
	Let the MSE of  any estimator $\hat{\thetavec}\in{\mathbb{R}}^M$
		be given by
	\be
	\label{MSE_def}
	{\text{MSE}}(\hat{\thetavec},\thetavec,\Lambda)\define
{\rm{E}}_{\thetavecsmall_\Lambda}\left[(\hat{\thetavec}-\thetavec)(\hat{\thetavec}-\thetavec)^T\right].
\ee
The following proposition states
 the relation between the MSE from \eqref{MSE_def}
and the MSSE  from \eqref{MSSE} for coherent estimators.
\begin{proposition}
\label{prop_relation}
For any  coherent estimator with the selective bias $	\bvec_k(\thetavec,\Lambda)$, as defined in \eqref{11_bias}, the MSE satisfies
	\beqna
	\label{MSE}
	{\text{\normalfont{MSE}}}(\hat{\thetavec},\thetavec,\Lambda)
		\hspace{6cm}\nonumber\\=
		{\rm{E}}_{\thetavecsmall_\Lambda}\left[\Cmat(\hat{\thetavec},\hat{\Lambda},\thetavec)\right] 	
		+\sum_{k=1}^K\pi_{k}(\thetavec_\Lambda){\thetavec}_{\Lambda_k^c}\zp({\thetavec}_{\Lambda_k^c}\zp)^T
		\hspace{1cm}
		\nonumber\\
			-\sum_{k=1}^K\pi_{k}(\thetavec_\Lambda)\left(\thetavec_{ \Lambda_k^c}\zp \bvec_k^T(\thetavec,\Lambda)
		+\bvec_k(\thetavec,\Lambda)(\thetavec_{ \Lambda_k^c}\zp)^T\right).
	\eeqna
			\end{proposition}
	\begin{IEEEproof} The proof appears in Appendix \ref{MSE_der_app}.
	\end{IEEEproof}
		
That is, 
	the  MSE in (\ref{MSE})  is the sum of the MSSE,
	the influence of the selective bias of the estimator, and an additional term,
	$\sum_{k=1}^K\pi_{k}(\thetavec_\Lambda){\thetavec}_{\Lambda_k^c}\zp({\thetavec}_{\Lambda_k^c}\zp)^T$,
		which is
	only a function of the selection rule and is not affected by the estimator, $\hat{\thetavec}$. Therefore,  by deriving a CRB-type lower bound on the MSSE we readily obtain  a lower bound on the MSE of any coherent estimator with a known selective bias function by using the relation in Proposition \ref{prop_relation}.

	By using the definition in \eqref{11_bias}, it can be verified that the selective bias can have a nonzero value only on  $ \Lambda_k$, and, thus, it has no overlap with  $\thetavec_{ \Lambda_k^c}$. Therefore, the diagonal elements of 
	$\thetavec_{ \Lambda_k^c}\zp \bvec_k^T(\thetavec,\Lambda)$ satisfy
	\be
	\label{trace1}
\left[\thetavec_{ \Lambda_k^c}\zp \bvec_k^T(\thetavec,\Lambda)\right]_{m,m}
	=0,~ m=1,\ldots,M,
	\ee
	and the trace of the matrix $\thetavec_{ \Lambda_k^c}\zp \bvec_k^T(\thetavec,\Lambda)$ is zero.
	By applying the trace operator on \eqref{MSE} and then 
	substituting \eqref{trace1}, we obtain 
			\beqna
	\label{SSE2_null2}
	{\text{Tr}}\left({\text{MSE}}(\hat{\thetavec},\thetavec,\Lambda)\right)= 
		{\text{Tr}}\left({\rm{E}}_{\thetavecsmall_\Lambda}[\Cmat(\hat{\thetavec},\hat{\Lambda},\thetavec_{\Lambda})]\right)\nonumber \\
	+\sum_{k=1}^K\pi_{k}(\thetavec_\Lambda){\text{Tr}}\left({\thetavec}_{\Lambda_k^c}\zp({\thetavec}_{\Lambda_k^c}\zp)^T\right),
	\eeqna
	where the last term in \eqref{SSE2_null2} is only a function of the true parameters, $\thetavec_{\Lambda}$,
	since it is assumed that $|\Lambda_k| \leq  M$ for any candidate model, $k=1,\ldots,K$. 
Similarly, by substituting \eqref{MSSE_marg2} in the  $m$th diagonal element of the MSE from \eqref{MSE} and using \eqref{SSE2_null2}, we obtain that 
	the marginal MSE on a specific parameter that belongs to the true model, $\theta_m$, is given by 
	\beqna
		\label{MSE_marg2}
		{\rm{E}}_{\thetavecsmall_\Lambda}\left[(\hat{\theta}_m-\theta_m)^2\right]=\hspace{4.75cm}
			\nonumber\\
		p_m(\thetavec_{\Lambda}) 
		{\rm{E}}_{\thetavecsmall_\Lambda}\left[(\hat{\theta}_{m}-\theta_{m})^2|m\in\hat{\Lambda}\right]
		+\left(1-p_m(\thetavec_{\Lambda})\right)\theta_m^2,
	\eeqna
	$\forall m\in{\Lambda}$,
 Since $\theta_m=0$, $\forall m\in{\Lambda^c}$, then, 	the marginal MSE on a null parameter is 
		\beqna
		\label{MSE_marg2_null}
		{\rm{E}}_{\thetavecsmall_\Lambda}\left[(\hat{\theta}_m-\theta_m)^2\right]
		=p_m(\thetavec_{\Lambda}) 
		{\rm{E}}_{\thetavecsmall_\Lambda}\left[\hat{\theta}_{m}^2|m\in\hat{\Lambda}\right],
	\eeqna
	$\forall m\in{\Lambda^c}$.

%%%%%%%%%%%%%%%%%%%%%%%%%%%%%
	\subsection{Selective CRB}
	\label{sCRBsec}
	Obtaining the estimator with the minimum MSSE among all coherent  estimators  with a specific selective bias is usually intractable. Thus, lower bounds on the MSSE and MSE of any coherent  estimator are useful for performance analysis and system design. In the following, a novel CRB for estimation after model selection, named here selective CRB, is derived. 
	To this end, we define the following post-model-selection likelihood gradient vectors:
	\be
	\label{l_define}
		\upsilonvec_k(\xvec,\thetavec_{\Lambda})\define \nabla_{\thetavecsmall_{\Lambda}}^T\log f(\xvec|{\hat{\Lambda}=\Lambda}_k;\thetavec_{\Lambda}),~~~\forall \xvec \in {\mathcal{A}}_k,
	\ee
	 	 for any $k=1,\ldots,K$ such that $\pi_k(\thetavec_{\Lambda})\neq 0$.
	The vectors $\upsilonvec_k(\xvec,\thetavec_{\Lambda})$ are all $|\Lambda|$-dimensional vectors,
	since the gradient is always w.r.t. the true parameter vector, $\thetavec_\Lambda$.
	%i.e. they have the same dimension as $\thetavec_\Lambda$ for any $k=1,\ldots,K$. 
	In addition, by substituting \eqref{bayes}	in \eqref{l_define} it can be verified that
		\be
	\label{l_define2}
		\upsilonvec_k(\xvec,\thetavec_{\Lambda})=\nabla_{\thetavecsmall_{\Lambda}}^T\log f(\xvec;\thetavec_\Lambda)-\nabla_{\thetavecsmall_{\Lambda}}^T\log \pi_k(\thetavec_{\Lambda}),		
	\ee
	$\forall \xvec \in {\mathcal{A}}_k$. 
	The marginal selective FIM is defined as
	\beqna
	\label{JJJdef}
		\Jmat_k(\thetavec_\Lambda)
		\define {\rm{E}}_{\thetavecsmall_\Lambda}[\upsilonvec_k(\xvec,\thetavec_{\Lambda})
		\upsilonvec_k^T(\xvec,\thetavec_{\Lambda})|{\hat{\Lambda}=\Lambda}_k],
	\eeqna
 	$k=1,\ldots,K$. 
 	For any $k=1,\ldots,K$, the  	$\Dmat_{k}(\Lambda)$ is a $M\times |\Lambda|$  matrix   with the  elements
	\beqna
	\label{Imat}
		\left[\Dmat_{k}(\Lambda)\right]_{m,l}
		 \define\left\lbrace		\begin{array}{cc}	
		1,&m\in{\Lambda}_k,~m=[\Lambda]_l\\
		0,&{\text{otherwise}}
		\end{array}\right.  ,
	\eeqna
	$\forall l=1,\ldots,M$,
 		where $[\Lambda]_l$ denotes the $l$th element of the true support set (not to be confused with the $l$th candidate support set, $\Lambda_l$).
 		Finally, the zero-padded gradient of the $k$th selective bias from \eqref{11_bias} is defined as
	\beqna
	\label{der_bias}
\Gmat_k(\thetavec,\Lambda)
 \define \nabla_{\thetavecsmall_{\Lambda}}\bvec_k(\thetavec,\Lambda),~\forall k=1,\ldots,K.
\eeqna
	We assume the following regularity conditions:
	\renewcommand{\theenumi}{C.\arabic{enumi}}
	\begin{enumerate}
		\item
		\label{cond1}
		The post-model-selection likelihood gradient vectors, $\upsilonvec_k(\xvec,\thetavec_{\Lambda})$, $k=1,\ldots,K$,	exist, and the selective FIMs,
		$\Jmat_k(\thetavec_\Lambda)$, $k=1,\ldots,K$, are well-defined and nonsingular matrices, $\forall \thetavec_\Lambda\in {\mathbb{R}}^{|\Lambda|}$. 
		\item
		\label{cond2}  
		The operations of integration w.r.t. $\xvec$ and differentiation w.r.t. $\thetavec_{\Lambda}$ can be interchanged, as follows:
		\beqna
			\nabla_{\thetavecsmall_{\Lambda}}\int_{\mathcal{A}_k} g(\xvec)  f(\xvec|{\hat{\Lambda}=\Lambda}_k;\thetavec_{\Lambda})\ud\xvec \hspace{2cm}
			\nonumber\\
			=\int_{\mathcal{A}_k} g(\xvec) \nabla_{\thetavecsmall_{\Lambda}} f(\xvec|{\hat{\Lambda}=\Lambda}_k;\thetavec_{\Lambda})\ud\xvec,
		\eeqna
		 $\forall k=1,\ldots,K$, for any measurable function,
		 $g:
		 \Omega_{\xvec}\rightarrow{\mathbb{R}}$,
		   $\forall \thetavec_\Lambda\in {\mathbb{R}}^{|\Lambda|}$. 		 Similar to the regularity conditions of the CRB (see, e.g. Chapter 2 in \cite{Theory_Statistics_book}),
		 	a sufficient condition for this to hold is
that the subset of $ \Omega_\xvec$ given by $ \{\xvec\in \Omega_\xvec|
f(\xvec|{\hat{\Lambda}=\Lambda}_k;\thetavec_{\Lambda})>0\}$  
is the same  $\forall \thetavec_\Lambda$.
	\end{enumerate} 
	\renewcommand{\theenumi}{\arabic{enumi}}
	The following theorem presents the proposed selective CRB.
	\begin{Theorem}
	\label{Th1}
		Let the regularity Conditions \ref{cond1}-\ref{cond2} be satisfied, $\hat{\Lambda}$ be a given selection rule,  and
		$\hat{\thetavec}$ be a coherent  estimator 	of $\thetavec$ with the support set, $\Lambda$,
				 where the selective bias of the estimator is $	\bvec_k(\thetavec,\Lambda)$, as defined in \eqref{11_bias}.
 Then, the MSSE satisfies
	\be
	\label{CRB}
		{\rm{E}}_{\thetavecsmall_{\Lambda}}[\Cmat(\hat{\thetavec},\hat{\Lambda},\thetavec)]\succeq \Bmat_{\text{sCRB}}(\thetavec_{\Lambda}),
	\ee
	where the selective CRB is given by
	\beqna
	\label{Rdef}
	\Bmat_{\text{sCRB}}(\thetavec_{\Lambda})\hspace{6cm}
	\nonumber\\
	\define\sum_{k=1}^K\pi_k(\thetavec_{\Lambda})\left(	\Psi_k(\thetavec,\Lambda)
+	\bvec_k(\thetavec,\Lambda)\bvec_k^T(\thetavec,\Lambda)\right),
	\eeqna
	in which 
	\beqna\label{psi_def}
	\Psi_k(\thetavec,\Lambda)\define\hspace{6cm}
	\nonumber\\
	(\Dmat_{k}(\Lambda)
+\Gmat_k(\thetavec,\Lambda))
	\Jmat_k^{-1}(\thetavec_{\Lambda})
(\Dmat_{k}(\Lambda)
+\Gmat_k(\thetavec,\Lambda))^T,
	\eeqna
	$\Dmat_{k}(\Lambda)$ is the zero-one  matrix defined in \eqref{Imat},
	and $\Jmat_k(\thetavec_{\Lambda})$ is the $k$th selective FIM, defined in (\ref{JJJdef}).
	Furthermore, the MSE from (\ref{MSE}) is bounded by
	\beqna
	\label{bound3_new}
		{\text{\em{MSE}}}(\hat{\thetavec},\thetavec_{\Lambda})
		\succeq\Bmat_{\text{sCRB}}(\thetavec_{\Lambda}) 	+\sum_{k=1}^K\pi_{k}(\thetavec_\Lambda){\thetavec}_{\Lambda_k^c}\zp({\thetavec}_{\Lambda_k^c}\zp)^T
			\nonumber\\
			-\sum_{k=1}^K\pi_{k}(\thetavec_\Lambda)\left(\thetavec_{ \Lambda_k^c}\zp\bvec_k^T(\thetavec,\Lambda)
		+\bvec_k(\thetavec,\Lambda)(\thetavec_{ \Lambda_k^c}\zp)^T\right).
	\eeqna
	\end{Theorem}
	\begin{IEEEproof} The proof appears in Appendix \ref{CRBProof}.
	\end{IEEEproof}
%%%%%%%%%%%%%%%%%%%%%%%%%%%%%%%%%%%%%%%%%%%%%%%%%%%%%%%%%%%%%%%%%%%%%%%%%%%%%%%%%%%%%%%%%%%%%%%%%%%%%%%%%%%%%%%

The MSE and MSSE bounds in Theorem \ref{Th1} are matrix bounds. As such, they imply  the associated marginal bounds on the diagonal elements and on the trace.
That is, by using the $m$th element of the MSSE from \eqref{MSSE_marg2}
		and the bound from \eqref{CRB}-\eqref{Rdef},
		we obtain 
	the marginal selective CRB on the MSSE of the $m$th element of $\thetavec$:
		\beqna
		\label{sCRBmarginalMSSE}
				p_m(\thetavec_{\Lambda}){\rm{E}}_{\thetavecsmall_\Lambda}\left[(\hat{\theta}_m-\theta_m)^2|m\in\hat{\Lambda}\right]
			\geq		
			[		\Bmat_{\text{sCRB}}(\thetavec_{\Lambda})]_{m,m}
			\nonumber\\
			=		
			\sum_{k\in\kappa_m}\pi_k(\thetavec_{\Lambda})
			\left(\left[\Psi_k(\thetavec,\Lambda)\right]_{m,m}
				+	\left[\bvec_k(\thetavec,\Lambda)\right]_{m,m}^2\right),
		\eeqna
		$\forall m=1,\ldots,M$,
		where $\kappa_m$ is defined in \eqref{kappa1}.
		Similarly, 
	 using the $m$th marginal MSE from \eqref{MSE_marg2}
and
		the matrix MSE bound from \eqref{bound3_new}  implies the
		following marginal MSE bounds:
		\beqna
		\label{sCRBmarginal}
			{\rm{E}}_{\thetavecsmall_{\Lambda}}\left[(\hat{\theta}_m-\theta_m)^2\right]\geq\hspace{4cm}\nonumber\\
			\sum_{k\in\kappa_m}\pi_k(\thetavec_{\Lambda})
			\left([\Psi_k(\thetavec,\Lambda)]_{m,m}
				+	[\bvec_k(\thetavec,\Lambda)]_{m,m}^2\right)
							\nonumber\\
			+\left(1-p_m(\thetavec_{\Lambda})\right)\theta_m^2,\hspace{4cm}
		\eeqna
		$\forall m =1,\ldots,M$, where $p_m(\thetavec_{\Lambda})$ is defined in \eqref{pm_def}.
				Summing 
				\eqref{sCRBmarginal}, over $m =1,\ldots,M$,
				and using the fact that $\thetavec_{\Lambda^c}=\zerovec$, we obtain the associated selective CRB on the  trace MSE:
		\beqna
		\label{sCRBtrace}
			{\text{Tr}}\left({\text{MSE}}(\hat{\thetavec},\thetavec_{\Lambda})\right)\hspace{5.5cm}
			\nonumber\\\geq
		\sum_{m=1}^{M}\sum_{k\in\kappa_m}\pi_k(\thetavec_{\Lambda})	\left([\Psi_k(\thetavec,\Lambda)]_{m,m}
				+	\left[\bvec_k(\thetavec,\Lambda)\right]_{m,m}^2\right)
			\nonumber\\+\sum\nolimits_{m\in\Lambda}\left(1-p_m(\thetavec_{\Lambda})\right)\theta_m^2.
			\hspace{3.5cm}
		\eeqna

%%%%%%%selective unbiased
In the following we present the selective CRB for selective unbiased estimators. By substituting \eqref{11_bias_zero} in
\eqref{psi_def}, we obtain that 
for selective unbiased estimators 
$\Psi_k(\thetavec,\Lambda)=\Dmat_{k}(\Lambda)
\Jmat_k^{-1}(\thetavec_{\Lambda})
\Dmat_{k}^T(\Lambda),
$ and, thus,  the selective CRB from \eqref{Rdef} is reduced to
\be
	\label{BsCRB}
		\Bmat_{\text{sCRB}}(\thetavec_{\Lambda})=\sum_{k=1}^K\pi_{k}(\thetavec_{\Lambda})\Dmat_{k}(\Lambda)\Jmat_k^{-1}(\thetavec_{\Lambda})\Dmat_{k}(\Lambda).
	\ee
Similarly, by substituting  \eqref{Imat} and \eqref{BsCRB} in \eqref{sCRBtrace},  the associated selective CRB on the  trace MSE of unbiased selective estimators is reduced to
		\beqna
		\label{sCRBtrace2}
			{\text{Tr}}\left({\text{MSE}}(\hat{\thetavec},\thetavec_{\Lambda})\right)
			\geq
		\sum_{l=1}^{|\Lambda|}\sum_{k\in\kappa_{[\Lambda]_l}}\pi_k(\thetavec_{\Lambda})[\Jmat_k^{-1}(\thetavec_{\Lambda})]_{l,l}
		\nonumber\\+\sum_{m\in\Lambda}\left(1-p_m(\thetavec_{\Lambda})\right)\theta_m^2,\hspace{1.75cm}
		\eeqna
		where $\kappa_{[\Lambda]_l}$ is defined in \eqref{kappa1}.
		
		It can be seen that for a selective unbiased estimator, as defined in \eqref{11_bias}, the selective CRB in \eqref{BsCRB} is  a lower bound only on the parameters that belong to the {\em{true}}  support set of the  vector, $\Lambda$,
		{\textcolor{black}{and does not include the null parameters.}}
		Thus, the influence of the selected  null parameters (i.e. 
 the parameters that  have been wrongly selected by the selection rule)
		on the estimation of the true parameters is via the selective bias and its gradient, from	\eqref{11_bias} and \eqref{der_bias}, respectively. 
	{\textcolor{black}{
		In some applications, it could be useful to take into account only
the estimation errors over 	the intersection of the true and estimated support, i.e.  only the values of $\hat{\thetavec}_{\Lambda\cap\hat{\Lambda}}\zp-{\thetavec}_{\Lambda\cap\hat{\Lambda}}\zp$. In these cases, one can use the unbiased selective CRB from \eqref{BsCRB} as the lower bound on the MSSE of the true parameters, as 
appears in our preliminary derivation of the  scalar selective CRB  in \cite{meirRoutSSP}.}}	
			
	Finally, the following Lemma presents  alternative formulations of the selective FIM that may be more tractable for some estimation problems.
	\begin{lemma}
		\label{lemma1}
		Assume that Conditions \ref{cond1}-\ref{cond2} are satisfied in addition to the following regularity conditions:
			\renewcommand{\theenumi}{C.\arabic{enumi}} 
	\begin{enumerate}
	 \setcounter{enumi}{2}
		\item
		\label{cond3}
		The second-order derivatives   of $f(\xvec|{\hat{\Lambda}=\Lambda}_k;\thetavec_{\Lambda})$  w.r.t. the elements of $\thetavec_\Lambda$ exist and are bounded and continuous  $\forall \xvec\in\mathcal{A}_k,~k=1,\ldots,K$.
		\item
		\label{cond4}
		The integral, $\int_{\mathcal{A}_k}  f(\xvec|{\hat{\Lambda}=\Lambda}_k;\thetavec_{\Lambda}){\ud}\xvec$, is twice differentiable  under the integral sign  w.r.t. the elements of $\thetavec_\Lambda$,  $ \forall k=1,\ldots,K$, $\thetavec_\Lambda\in {\mathbb{R}}^{|\Lambda|}$. 
	\end{enumerate} 
	\renewcommand{\theenumi}{\arabic{enumi}}
		Then, the $k$th  selective FIM in (\ref{JJJdef})
		satisfies
		\beqna
		\label{q}
		\Jmat_k(\thetavec_\Lambda)=-{\rm{E}}_{\thetavecsmall_\Lambda}\left[\left.\nabla_{\thetavecsmall_\Lambda}\nabla_{\thetavecsmall_\Lambda}^T \log f(\xvec;\thetavec_\Lambda) \right|{\hat{\Lambda}=\Lambda}_k\right]\hspace{0.3cm}\nonumber\\
		+\nabla_{\thetavecsmall_\Lambda}\nabla_{\thetavecsmall_\Lambda}^T \log\pi_{k}(\thetavec_{\Lambda}),\hspace{3cm}
		\\
			\label{emp_Jk}
		=		{\rm{E}}_{\thetavecsmall_\Lambda}\left[\left.
		\nabla_{\thetavecsmall_{\Lambda}}^T \log f(\xvec;\thetavec_{\Lambda})
		\nabla_{\thetavecsmall_{\Lambda}}\log f(\xvec;\thetavec_{\Lambda}) \right|{\hat{\Lambda}=\Lambda}_k
		\right]
		\nonumber\\
				-\left(\nabla_{\thetavecsmall_{\Lambda}} \log \pi_{k}\left(\thetavec_{\Lambda}\right)\right)^T\nabla_{\thetavecsmall_{\Lambda}} \log \pi_{k}\left(\thetavec_{\Lambda}\right),\hspace{1.5cm}
	\eeqna
		 $ \forall k=1,\ldots,K$, $\thetavec_\Lambda\in {\mathbb{R}}^{|\Lambda|}$.
	\end{lemma}
\begin{IEEEproof} The proof appears in Appendix \ref{proofLemma}.
 \end{IEEEproof}
	%%%%	

	\subsection{Special cases and relation with other CRB-type bounds}
	\label{specialCases} 	
	\subsubsection{Single model}
	When only a single model is assumed, i.e.  ${\hat{\Lambda}=\Lambda}_1=\Lambda$ and $\pi_1(\thetavec_{\Lambda})=1$ for any selection rule,  the SSE, selective unbiasedness, and selective CRB are reduced to the MSE, mean-unbiasedness, and CRB for estimating $\thetavec_{\Lambda}$. Thus, the proposed paradigm generalizes the conventional non-Bayesian parameter estimation.
	%%%%%%%%%%%%%%%
	\subsubsection{Nested models and the relation to SMS-CRB}
	A model class is nested  if smaller models are always special cases of larger models.
	Thus, in this special case
 we assume a model \textit{order} selection problem in which $\Lambda_1\subset\ldots \subset \Lambda_{k_t}\ldots\subset\Lambda_k$, $\forall k=1,\ldots,K$,
	where $k_t$ is the true model, i.e.  $\Lambda=\Lambda_{k_t}$.
	In this special case, the  matrix $\Dmat_{k}(\Lambda)$ from \eqref{Imat} is given by
	\be
	\label{Dmat_nested}
	\Dmat_{k}(\Lambda)=
	\left\{
		\begin{array}{lr}
	\begin{bmatrix}
	\Imat_{|\Lambda_k|} &\zerovec\\
	\zerovec&\zerovec
\end{bmatrix} & {\text{if }} 1\leq k\leq {k_t}\\
	\begin{bmatrix}
\Imat_{|\Lambda|}
\\
	\zerovec
\end{bmatrix}
& {\text{if }}{k_t} \leq k\leq K
\end{array}
\right. .
\ee
By substituting  \eqref{Dmat_nested} in the unbiased selective CRB   from  \eqref{BsCRB},
	we obtain
		\beqna
	\label{BsCRB_nested}
		\Bmat_{\text{sCRB}}(\thetavec_{\Lambda})=	\sum_{k=1}^{K}\pi_{k}(\thetavec_{\Lambda})
		\hspace{4cm}\nonumber\\\times
		\begin{bmatrix}
	[\Jmat_k^{-1}(\thetavec_\Lambda)]_{\{1:\min\{|\Lambda_k|,\Lambda\},
	1:\min\{|\Lambda_k|,\Lambda\}\}} &\zerovec\\
	\zerovec&\zerovec
\end{bmatrix},
	\eeqna
	$\forall  k=1,\ldots,K$.

		The SMS-CRB  bound  from \cite{Sando_Mitra_Stoica2002} was developed 
		for the problem of model order selection  with nested models
		under the assumptions:
		\renewcommand{\theenumi}{A.\arabic{enumi}} 
	\begin{enumerate}
		\item\label{A1} The order selection rule is such that asymptotically $\Pr({\hat{\Lambda}=\Lambda}_k;\thetavec_\Lambda)=0$,
		for any $k<k_t$.
	Hence,	only  overestimation of
the order is considered.
\item
\label{A2}  The FIMs under the $k$th candidate model,
defined as
	\be
	\label{oraclek_FIM}
	\tilde{\Jmat}_k(\thetavec_{\Lambda_k})\define -{\rm{E}}_{\thetavecsmall_{\Lambda_k}}[\nabla_{\thetavecsmall_{\Lambda_k}}(\nabla^T_{\thetavecsmall_{\Lambda_k}}\log f(\xvec;\thetavec_{\Lambda_k}))], 
	\ee
are nonsingular matrices for any $k=1,\ldots,K$.
\end{enumerate}
	\renewcommand{\theenumi}{\arabic{enumi}}
	Under Assumptions \ref{A1}-\ref{A2},
	the SMS-CRB is given by \cite{Sando_Mitra_Stoica2002}
\beqna
\label{SMS-CRB}
\Bmat_{\text{SMS-CRB}}(\thetavec)\define \sum_{k=k_t}^K\pi_k(\thetavec_\Lambda)
\Fmat_k(\thetavec_{\Lambda_k})\define\begin{bmatrix}
	\tilde{\Jmat}_k^{-1}(\thetavec_{\Lambda_k}) &\zerovec\\
	\zerovec&\zerovec
\end{bmatrix}.
\eeqna
It can be seen that  the proposed selective CRB for nested models from \eqref{BsCRB_nested}
 has a similar structure to the SMS-CRB  bound  from \eqref{SMS-CRB}.
However, 
the proposed selective CRB accounts for both overestimation and underestimation of
the model order,
 while the SMS-CRB accounts only for overestimation.
The selective CRB is based on a different selective FIM for each model, that takes into account the selection rule,
while the SMS-CRB
 is based on averaging over the FIMs of the different candidate models, as can be seen from comparing \eqref{BsCRB_nested} and \eqref{SMS-CRB}.
Finally, our bound is not limited to the nested  setting and is shown to be tighter than the SMS-CRB in simulations. 
	%%%%%%%%%%%%%%%%
	
	\subsubsection{Data-independent selection rule}
	\label{subsub_rand}
	In this degenerated case, we consider  a random selection rule, which is independent of the data  and of its parameters.
	Thus, 
	the derivative of the log of the probability of selection of the $k$th model w.r.t.  $\thetavec_{\Lambda}$ vanishes:
	\be
	\label{der_zero}
\nabla_{\thetavecsmall_\Lambda}^T \log\pi_{k}(\thetavec_{\Lambda})=\zerovec.
	\ee
	In addition,  since the selection is independent of the observation vector, $\xvec$, then
	\beqna
	\label{conditional_random}
	-{\rm{E}}_{\thetavecsmall_\Lambda}\left[\left.\nabla_{\thetavecsmall_\Lambda}\nabla_{\thetavecsmall_\Lambda}^T \log f(\xvec;\thetavec_\Lambda) \right|{\hat{\Lambda}=\Lambda}_k\right]=\Jmat(\thetavec_\Lambda),
	\eeqna
	$\forall k=1,\ldots,K$, where
	\beqna
	\Jmat(\thetavec_\Lambda)\define
	-{\rm{E}}_{\thetavecsmall_\Lambda}[\nabla_{\thetavecsmall_{\Lambda}}(\nabla^T_{\thetavecsmall_{\Lambda}}\log f(\xvec;\thetavec_\Lambda))]
	\eeqna
	is the oracle FIM, which {\textcolor{black}{is based on}} knowledge of the {\textcolor{black}{true}} model.
	By substituting (\ref{der_zero}) and (\ref{conditional_random})
	in the selective FIM from (\ref{q}), we obtain that for a random selection rule,
	\beqna
	\label{FIM}
		\Jmat_k(\thetavec_\Lambda)=\Jmat(\thetavec_\Lambda),~\forall k=1,\ldots,K.
	\eeqna
	Substitution of (\ref{FIM}) in  the selective CRB from \eqref{BsCRB}, results in
		\beqna
	\label{BsCRB_random}
		\Bmat_{\text{sCRB}}(\thetavec_{\Lambda})=
		\sum_{k=1}^K\pi_{k}(\thetavec_{\Lambda})\Dmat_{k}(\Lambda)\Jmat^{-1}(\thetavec_\Lambda)\Dmat_{k}(\Lambda).
	\eeqna
	%By substituting the definition of $\Dmat_{\Lambda\cap\Lambda_k}\zp$ from \eqref{Imat} in \eqref{BsCRB_random}
	%we obtain that the $(m,l)$ element of \eqref{BsCRB_random} is given by
			%\beqna
	%\label{BsCRB_random2}
		%\left[\Bmat_{\text{sCRB}}(\thetavec_{\Lambda})\right]_{m,l}\hspace{5cm}
		%\nonumber\\=
		%\left[(\Jmat(\thetavec_\Lambda))^{-1}\right]_{m,l}
		%\sum_{k=1}^K\pi_{k}(\thetavec_{\Lambda}){\mathbf{1}}_{\{m\in\Lambda_k\}}
		%{\mathbf{1}}_{\{l\in\Lambda_k\}}\hspace{0.8cm}
		%\nonumber\\
		%=\left[(\Jmat(\thetavec_\Lambda))^{-1}\right]_{m,l} \Pr\left(m,l{\text{ have been selected}};\thetavec_\Lambda\right).
	%\eeqna
	Thus, for a random selection rule,  the selective CRB is a weighted average over the elements of the oracle CRB, $\Jmat^{-1}(\thetavec_\Lambda)$,
	where the weights are defined by the probability of selection.
	%In particular, if Assumption \ref{A1} of the SMS-CRB holds, then 
%	\eqref{BsCRB_random} is reduced to 
%$
%		\Bmat_{\text{sCRB}}(\thetavec_{\Lambda})=
%	\begin{bmatrix}
%	\Jmat^{-1}(\thetavec_\Lambda) &\zerovec\\
%	\zerovec&\zerovec
%\end{bmatrix}.
%$
\subsubsection{Relation with the misspecified CRB}
Estimation after model selection can be interpreted as 	
 estimation with a misspecified (or mismatched) model 
	\cite{richmondHorowitz,vuong,pajovic,Fortunati_Gini_Greco2016,white1982maximum},
	where 
	the true pdf is $f(\xvec;\theta_{\Lambda})$ for any $\xvec \in \Omega_\xvec$, and
	the assumed pdf (which may be wrong)  is 
	\beqna
	\label{tilde_f}
	\tilde{f}(\xvec;\thetavec)={\text{const}}\times \sum_{k=1}^{K} f(\xvec;\theta_{\Lambda_k}) {\mathbf{1}}_{\{\xvec \in {\mathcal{A}}_k\}},
	\eeqna
	where ``${\text{const}}$" is the normalization factor of the pdf. 
The sets
	${\mathcal{A}}_k$ are defined in \eqref{Akk} and are associated by the predetermined selection rule, $\hat{\Lambda}$.
However, it should be noted that the misspecified CRB in 	\cite{richmondHorowitz,vuong,pajovic,Fortunati_Gini_Greco2016} for the considered post-model-selection scheme is a lower bound on 
$ {\rm{E}}_{\thetavecsmall_\Lambda}[
(\hat{\thetavec}-\muvec)
(
\hat{\thetavec}-\muvec)^T]$, where
$
\muvec\define {\rm{E}}_{\thetavecsmall_\Lambda}[\hat{\thetavec}]
$.
%If we impose the coherency of the estimators, as defined in Definition \ref{cohdef}, then
%$\muvec=
%{\rm{E}}_{\thetavecsmall_\Lambda}[\hat{\thetavec}_{\hat{\Lambda}}\zp]$
%and  $ {\rm{E}}_{\thetavecsmall_\Lambda}[
%(\hat{\thetavec}-\muvec)
%(
%\hat{\thetavec}-\muvec)^T]$
This risk function can be interpreted as the squared-error  of the average selected parameter vector or the estimator, which is different from the SSE cost function in \eqref{aSSE}. The SSE is more appropriate for evaluating post-model-selection estimation error that aims to be close to the true parameter vector, $\thetavec_{\Lambda}$. In addition, the misspecified CRB does not take into account the coherency of the estimators, as defined in Definition \ref{cohdef}, which is a significant aspect of the considered scheme.
These differences are since, in contrast to  the misspecified CRB, we consider a well-specified architecture in which  the full  set of candidate models is  known. 
%%%%%%%%%

\subsection{Practical implementation of the selective CRB}
	\label{practical_sec}
	The proposed selective CRB  from Theorem \ref{Th1},
	as well as its different versions described in \eqref{sCRBmarginalMSSE}-\eqref{sCRBtrace2},
	requires the calculation of the selection probability, $\pi_k(\thetavec_{\Lambda})$,
	the marginal selective FIM, 
	$\Jmat_k(\thetavec_{\Lambda})$,
	and the  gradient of the $k$th selective bias, $\Gmat_k(\thetavec,\Lambda)$,
	from \eqref{pi_k},  \eqref{JJJdef}, and \eqref{der_bias},
respectively,
	for each
model. 
When the number of models
increases,  the efficient evaluation of the probabilities of selection,  the selective FIMs, and the bias gradients, is hard.
One way to reduce the computational complexity of 
the proposed bound is
by selecting a subset of models and replacing the sum in \eqref{Rdef} or \eqref{BsCRB} by a sum over the subset.
This approach is similar to the one used in the Barankin bound \cite{BARANKIN}, in which a
set of arbitrary test points is used to compute the bound.
The resultant
 bound is still a valid lower bound,
since  we only removed non-negative terms that are associated with the neglected models. 
In order to reduce the set of models and simultaneously to increase the tightness, 
it intuitively seems more efficient to use the  models with the highest probability of selection.\\
Even after reducing the number of candidate models,
 the proposed bound may be intractable.
In this case, the selective CRB can be approximated by  low-complexity  methods.
Similarly to the empirical FIM in \cite{berisha2015empirical,spall2005monte},  
a Monte Carlo approach can be developed to approximate the selective FIMs and the probability of selection by using the stochastic approximation family of algorithms,  which results in the empirical selective CRB. 
The multidimensional integrals needed to calculate the bound are obtained by 
drawing samples directly from the  true distribution,   $f\left(\xvec;\thetavec_{\Lambda}\right)$, for a given $\thetavec_{\Lambda}$, as described, for example, in \cite[Ch.~2]{robert2013monte}, or by using Markov chain Monte Carlo (MCMC) samplers \cite[Ch.~6]{robert2013monte}.
 For example,
 the probability of selecting the $k$th model from \eqref{pi_k} can be written as
\beqna
\label{emp_pik}
	\pi_{k}\left(\thetavec_{\Lambda}\right)=\int_{\Omega_\xvec}f\left(\xvec;\thetavec_{\Lambda}\right)\mathbf{1}_{\{\xvec\in\mathcal{A}_k\}}\ud\xvec.
\eeqna
Thus, we perform the data generation step as follows: 
we draw $D$ independent and identically distributed (i.i.d.) samples, $\{\tilde{\xvec}^{(d)}\}_{d=1}^D$, from the true pdf, $f(\xvec;\thetavec_{\Lambda})$. We use these samples in order to approximate \eqref{emp_pik}:
\beqna
\label{emp_pik_app}
\pi_{k}\left(\thetavec_{\Lambda}\right)\approx\frac{1}{D}\sum_{d=1}^{D}\mathbf{1}_{\{\tilde{\xvec}^{(d)}\in\mathcal{A}_k\}}.
\eeqna
From the strong law of large numbers, the approximation in \eqref{emp_pik_app} converges almost surely to the probability of selection in \eqref{emp_pik}.
Similar  approximations can be used for the selective FIM, where the specific structure of the post-model-selection log-likelihood in the selective FIM in its version in \eqref{emp_Jk} makes the calculation tractable, in a similar manner to our derivation in Section IV in \cite{Harel_Routtenberg_2019} for estimation after
 {\em{parameter}} selection.
Then, by replacing  the selective FIM and the probability of selection  with these approximations in the selective CRB expression  from \eqref{Rdef} or \eqref{BsCRB}, one can obtain the empirical selective CRB.
Due to space limitations, the full details of the empirical selective CRB are not presented in this paper.

%%%%%%%%%%%%%%%%%

%It should be noted that the specification of the biased CRB requires an {\em{a-priori}} choice of the bias gradient. Moreover, the selective bias  may be intractable.  
	%%%%%%%%%%%%%%%%%%%5

	%%%%%%%%%%%%%%%%%%%%%%%%%%%%%%%%%%%%%%%%%%%%%%%%%%%%%%%%%%%%%%%%%%%%%%%%%%%%%%%%%%%%%%%%%%%%%%%%%%%%%%%%%%%
	\section{Sparse vector estimation}
	\label{sparse}
In this section, we derive the selective CRB for the special case  of estimating an unknown sparse  vector, $\thetavec$, from noisy linear observations. This problem can be formulated as
\be
\label{sparseMeas}
\xvec = \Amat \thetavec+\wvec=\Amat_\Lambda\thetavec_\Lambda+\wvec,
\ee
where $\Amat\define[\avec_1,\ldots,\avec_M]\in{\mathbb{R}}^{L\times M}$ is a known measurement matrix, whose columns  satisfy $\avec_m^T\avec_m=1$,  $\forall m=1,\ldots, M$, and $\wvec \in{\mathbb{R}}^{L}$ is an independent noise vector. 
It is assumed that $\Amat_\Lambda$ is a full-rank matrix and that $|\Lambda|\ll M$, i.e. only a small number of elements in the unknown parameter vector, $\thetavec$, may be nonzero,
where the exact sparsity level (size of the support set)  is unknown.
%We aim to estimate 
%both the support set, $\Lambda$, and the 
%values of the unknown parameter vector,  
%$\thetavec_\Lambda$.
The candidate models include the different possible support sets of $\thetavec$,
$\Lambda_k$, $k=1,\ldots,K$.

Solving model-selection procedures for sparse vector estimation is known to be an NP-hard problem. 
Thus, we assume the
simple selection criterion of one step thresholding (OST) (see, e.g. \cite{donoho1994ideal,slavakis2013generalized,WassermanOST,OST}), which states that 
\be
\label{sparseSelection}
m\in \hat{\Lambda}\text{ if }|\avec_m^T\xvec|>c,~\forall m=1,\ldots,M,
\ee
where $c$ is a positive, user-selected threshold. This rule simply correlates the observed signal with all the frame vectors and selects the indices where the correlation energy exceeds a certain level, $c$.
For the sake of simplicity, we develop the selective CRB for the common setup of additive Gaussian noise. That is, the noise vector, $\wvec$ from (\ref{sparseMeas}), is assumed to be an i.i.d. zero-mean vector with covariance matrix $\sigma^2\Imat_L$, where $\sigma^2$ is known. We denote the standard normal pdf and cumulative distribution function (cdf) as
$
\phi(x)\define\frac{1}{\sqrt{2\pi}}{\rm{e}}^{-\frac{x^2}{2}}$
 and
$
\Phi(x)\define\int_{-\infty}^{x}\phi(t){\ud}t$,
respectively. 
In addition, we use the notations
\beqna
\label{alpha_def}
\alpha_m\define \frac{c-\avec_m^T\Amat_{\Lambda}\thetavec_{\Lambda}}{\sigma},~m=1,\ldots,M
\eeqna
and
\beqna
\label{bbb}
\beta_m\define \frac{-c-\avec_m^T\Amat_{\Lambda}\thetavec_{\Lambda}}{\sigma}~m=1,\ldots,M.
\eeqna
%%%%%%%%%%%%%%%%%%%%%%%%%%%%%
\begin{Theorem}
\label{Th2}
	The selective FIM from \eqref{JJJdef}  for the  model in (\ref{sparseMeas}), 
	where $\wvec$ is a zero-mean Gaussian vector with a covariance matrix $\sigma^2\Imat_L$
	and with the OST selection rule, is given by
	\beqna
	\label{spsJ}
	\Jmat_k(\thetavec_\Lambda)
	%=\frac{1}{\sigma^2}\Amat_\Lambda^T\left(\Imat_L+\Qmat_k\right)\Amat_\Lambda
	=\Jmat(\thetavec_\Lambda)+
	\frac{1}{\sigma^2}\Amat_\Lambda^T\Qmat_k\Amat_\Lambda,
	\eeqna
where
\beqna
\label{oracle_J}
\Jmat(\thetavec_\Lambda)=
\frac{1}{\sigma^2}\Amat_\Lambda^T\Amat_\Lambda
\eeqna
 is the oracle FIM for this case, which {\textcolor{black}{is calculated for}} the true support set of the sparse vector, $\Lambda$,
and
	\beqna
\label{Q_def}
\Qmat_k\define
\sum_{l\in\Lambda_k}
{\avec_l\avec_l^T}\left\{
\frac{\phi(\alpha_l)\alpha_l
-\phi(\beta_l)\beta_l}
{1-\Phi(\alpha_l)+\Phi(\beta_l)}\right.\hspace{1.5cm}
\nonumber\\\left.
-
\frac{\left(\phi(\alpha_l)
-\phi(\beta_l)\right)^2}
{\left(1-\Phi(\alpha_l)+\Phi(\beta_l)\right)^2}\right\}\hspace{2.25cm}
\nonumber\\
+\sum_{m\notin\Lambda_k}{\avec_m\avec_m^T}\left\{
\frac{-\phi(\alpha_m)\alpha_m
+\phi(\beta_m)\beta_m}{\Phi(\alpha_m)-\Phi(\beta_m)}\right.
\nonumber\\\left.
-
\frac{\left(\phi(\alpha_m)
-\phi(\beta_m)\right)^2}{\left(\Phi(\alpha_m)-\Phi(\beta_m)\right)^2}\right\},\hspace{2cm}
\eeqna
for any $k=1,\ldots,K$.
\end{Theorem}
\begin{IEEEproof} The proof appears in Appendix \ref{sparseAPP}.
	\end{IEEEproof}

It can be seen that asymptotically, i.e. when $\sigma\rightarrow\infty$, 
the elements of the matrix  $\Qmat_k$ from \eqref{Q_def} converge to zero, $\forall k=1,\ldots,K$,
faster than those of the oracle FIM from \eqref{oracle_J}.
Thus, asymptotically, the selective FIM from
	\eqref{spsJ} converges to the oracle FIM from \eqref{oracle_J}.
As a result, in the asymptotic, small-error region  the proposed selective CRB and the oracle CRB coincide. 
However, in the large-error region, while the oracle CRB is known to be non-informative, the  selective CRB is an informative lower bound, as shown in the simulations,  since it takes into account the probability that the OST selection rule  selects wrong parameters and/or misses true parameters.

The selective FIM from Theorem \ref{Th2} can be used to compute the different versions of the selective CRB
from \eqref{Rdef} and \eqref{bound3_new}-\eqref{sCRBtrace},
under the assumption that 
$\Amat_\Lambda^T\Amat_\Lambda+\Amat_\Lambda^T\Qmat_k\Amat_\Lambda$, $k=1,\ldots,K$, are nonsingular matrices. 
In particular, by substituting \eqref{foldedCDF} from  Appendix \ref{sparseAPP} 
 and \eqref{spsJ}  in \eqref{sCRBtrace2} we obtain that the selective CRB  on the  trace MSE  of selective unbiased estimators is given by
\beqna
		\label{sCRBtracesparce}
			{\text{Tr}}\left({\text{MSE}}(\hat{\thetavec},\thetavec_{\Lambda})\right)\hspace{5cm}
			\nonumber\\\geq
		\sigma^2
		\sum_{l=1}^{|\Lambda|}\sum_{k\in\kappa_{[\Lambda]_l}}
		 \pi_{k}(\thetavec_{\Lambda})
		\left[\left(\Amat_\Lambda^T\left(\Imat+\Qmat_k\right)\Amat_\Lambda\right)^{-1}\right]_{	l,l}
		\nonumber\\
		+\sum_{m\in\Lambda}(\Phi(\alpha_m)-\Phi(\beta_m))\theta_m^2,\hspace{2.5cm}
		\eeqna
	where
	$\Qmat_k$ is defined in \eqref{Q_def}, and,
	 according to Appendix \ref{sparseAPP}: 
	\beqna
	\label{prob_sparse}
\pi_k(\thetavec_{\Lambda})=\hspace{6cm}
\nonumber\\
\prod_{l\in\Lambda_k}
\left(1-\Phi(\alpha_l)+\Phi(\beta_l)\right) \prod_{m\notin\Lambda_k}\Phi(\alpha_m)-\Phi(\beta_m),
\eeqna
 $\forall k=1,\ldots,K$.
 Since in many sparse estimation scenarios one cannot construct any estimator which is unbiased for all sparsely representable parameters \cite{sparse_con}, the biased selective CRB from Theorem \ref{Th1} should be used for cases where the bias gradient of the considered estimator is tractable.

	According to Subsection \ref{practical_sec}, 
	we can reduce the computational complexity of 
the bound from \eqref{sCRBtracesparce}-\eqref{prob_sparse}
by selecting a subset of models such that $k\in\kappa_m$, $m\in\Lambda$.
It should be noted that the SMS-CRB from \eqref{SMS-CRB}
		cannot be computed for the sparse setting with $L<M$ since the SMS-CRB requires that
		the FIMs under the $k$th candidate model from \eqref{oraclek_FIM} will be
 nonsingular matrices for any $k=1,\ldots,K$, while in the sparse setting these are usually not full-rank matrices.

	%%%%%%%%%%%%%%%%%%%%%%%%%%%%%%%%%%%%%%%%%%%%%%%%%%%%%%%%%%%%%%%%%%%%%%%%%%%%%%%%%%%%%%%%%%%%%%%%%%%
\section{Examples}
	\label{simulations}
	In this section, the proposed selective CRB is evaluated  
	and its performance is compared with that of the   SMS-CRB  \cite{Sando_Mitra_Stoica2002} from \eqref{SMS-CRB},  the oracle CRB, and with  the performance of the maximum selected likelihood (MSL) estimator.
	The MSL estimator
is obtained by first choosing a model based on a selection rule
and
then 
	maximizing the  likelihood  of the selected model.
	The  performance of the MSL estimator  is
	evaluated using   $20,000-100,000$ Monte-Carlo simulations.
	Unless otherwise noted,
	in the following simulations we present the unbiased  selective CRB from \eqref{sCRBtrace2}.
	Finally, it should be noted that in all simulations the probability of selection is analytically computed.

	%%%%%%%%%%%%%%%%%%%%%%%%%%%%%%%%%%%%%%%%%%%%%%%%%%%%%%%%%%%%%%%%%%%%%%%%%%%%%%%%%%%%%%%%%%%%%%%%%%%%%%%%%%%%%%%%%%%%%%%%
	\subsection{Example 1 - General linear model}
	The general linear model  is applied to a large set of problems in different
	fields of science and engineering \cite{graybill1976theory,djuric1998asymptotic,Wiesel_Eldar_Yeredor2008}.
	Under this model and the $k$th  candidate model, the observations obey
	\be
	\label{AICmodel}
\xvec=\Hmat_{k}\thetavec_{\Lambda_k}+\wvec,
	\ee
	where 
	 $\xvec\in{\mathbb{R}}^{N}$ is  the observed vector, 
	the matrices $\Hmat_{k}\in\mathbb{R}^{N\times|\Lambda_{k}|}$, $k=1,\ldots,K$, are assumed to be known full column rank matrices,
	$\thetavec\in{\mathbb{R}}^{M}$ is a  deterministic unknown vector,
and $\wvec$ is a zero-mean Gaussian random vector with mutually
independent elements, each with known variance $\sigma^2$.
	
	The coherent MSL estimator, under the assumption that the $k$th model has been selected,
	is given by (p. 186 in \cite{Kay_estimation})
	\be
	\label{ML_reg}
	\hat{\thetavec}^{{\text{ML}}|k}_{\Lambda_k}=\left({\Hmat}_{k}^T{\Hmat}_{k}\right)^{-1}{\Hmat}_{k}^T\xvec,~\forall k=1,\ldots,K,
	\ee
  where the other parameters of the MSL estimator are set to zero, 
i.e. 		$\hat{\thetavec}^{{\text{ML}}|k}_{\Lambda_k^c}=\zerovec$,
as   in  \eqref{coherency_gen}.
 The notation $|k$ denotes that the $k$th model was used
for the estimation.

We assume here selection rules from the GIC  family \cite{stoica2004model}.
For the considered model, the GIC 
 is given by
\beqna
\hat{\Lambda}^{\rm{GIC}}(\xvec)=\arg\min_{\Lambda_1,\ldots,\Lambda_K}	{\rm{GIC}}(\Lambda_k,N),
\eeqna
where
	\beqna
	\label{GIC0}
{\rm{GIC}}(\Lambda_k,N)\define-2\log{f(\xvec;\hat{\thetavec}^{{\text{ML}}|k})}+\tau\left(N,|\Lambda_k|\right)|\Lambda_k|
	\eeqna
	and 
	$\tau(N,|\Lambda_k|)$ is a penalty term. In particular,
for the two  widely-used AIC and MDL criteria we have 
\be
\tau\left(N,|\Lambda_k|\right)=\left\{\begin{array}{ll}2&{\text{AIC}}\\
\log N & {\text{MDL}}
\end{array}\right. .
\ee
	By substituting the model from \eqref{AICmodel}
	and the $k$th MSL estimator from \eqref{ML_reg} into \eqref{GIC0}, and removing constant terms,
	we obtain
		\beqna
	\label{GIC}
{\rm{GIC}}(\Lambda_k,N)
=\frac{1}{\sigma^2}||\Pmat_{\Hmat_{k}}^\bot\xvec||_2^2+\tau\left(N,|\Lambda_k|\right)|\Lambda_k|.
	\eeqna

For this model, it can be shown
that \cite{Kay_estimation}
\beqna
\label{FIM_GLM}
\nabla_{\thetavecsmall_\Lambda}\nabla_{\thetavecsmall_\Lambda}^T \log f(\xvec;\thetavec_\Lambda)
=-\frac{\tilde{\Hmat}^T{\tilde{\Hmat}}}{\sigma^2},
\eeqna
where $\tilde{\Hmat}$ is the true measurement matrix, $\tilde{\Hmat}\in\{\Hmat_k\}_{k=1}^K$.
By substituting \eqref{FIM_GLM} in (\ref{q}), it can be verified that,  where
the
correct model is $k$, the selective FIMs are
	\beqna
	\label{Jk_reg}
	\mathbf{J}_k(\thetavec)
	=\frac{\tilde{\Hmat}^T\tilde{\Hmat}}{\sigma^2}+
	\nabla_{\thetavecsmall_\Lambda}(\nabla^T_{\thetavecsmall_\Lambda}\log\pi_k(\thetavec_\Lambda)),~k=1,\ldots,K.
	\eeqna
%	$k=1,\ldots,K$.
%By substituting \eqref{Jk_reg} in the various versions of the selective CRB
%from \eqref{BsCRB}-\eqref{sCRBtrace},
%	we obtain the associated lower bounds for this case.
	
In general, the probability of selection, $\pi_k(\thetavec_\Lambda)$,  does not have
an analytical form.
For the sake of simplicity, in the following  simulations we set $K=2$,
$\Hmat_1=\hvec_1$, and 
$\Hmat_2=[\hvec_1,\hvec_2]$. 
Thus, $\Lambda_1=\{1\}$, $\Lambda_2=\{1,2\}$, $\thetavec_{\Lambda_1}=\theta_1$,   $\thetavec_{\Lambda_2}=\thetavec\in{\mathbb{R}}^2$.
We simulate data where
 $k=2$ is the true model, i.e. $\thetavec_{\Lambda}=\thetavec$.
 Thus, in this case there are no null parameters that have been wrongly selected, and the influence of the selection rule is via the coherency property from \eqref{coherency_gen}.
 This situation
 arises in many real-world scenarios with non-synthetic data and with no real null parameters. 
According to \eqref{GIC} and by using some algebraic manipulations, the probability of selecting  the $k=2$ model
 is
\beqna
\label{60}
\pi_2(\thetavec)=
\Pr\left(-\frac{1}{\sigma^2}\xvec^T(\Pmat_{\Hmat_2}^\bot-\Pmat_{\Hmat_1}^\bot)\xvec
\geq
\gamma;\thetavec\right)
\\
	\label{pi2_reg}
=Q_{\frac{1}{2}}(\sqrt{\lambda},\sqrt{ \gamma}),\hspace{3.65cm}
	\eeqna
	where 
		$\gamma \define  2\tau(N,2)-\tau(N,1)$,
		$\lambda=\frac{\theta_2^2}{\sigma^2} \frac{||\hvec_1||_2^2||\hvec_2||_2^2-(\hvec_1^T\hvec_2)^2}{||\hvec_1||_2^2}$,
and
	$Q_m(\cdot,\cdot)$ is the general Marcum Q-function of order $m$.
	The probability in \eqref{pi2_reg} is obtained by using the fact that
$-\frac{1}{\sigma^2}\xvec^T(\Pmat_{\Hmat_2}^\bot-\Pmat_{\Hmat_1}^\bot)\xvec
	$
	has a noncentral $\chi$-squared distribution with 1  degree of freedom and a non-centrality parameter $\lambda$ (see, e.g.
	\cite{5701798}).
	%and Chapter 2 in \cite{Kay_detection}
	Since 
	$\pi_2(\thetavec)$ is only a function of $\theta_2$, 
	the only nonzero element of $\nabla_{\thetavecsmall_\Lambda}(\nabla^T_{\thetavecsmall_\Lambda}\log\pi_2(\thetavec))$ from \eqref{Jk_reg}
	is its $(2,2)$th element, which satisfies
	\beqna
	\label{der22}
	\left[\nabla_{\thetavecsmall_\Lambda}(\nabla^T_{\thetavecsmall_\Lambda}\log\pi_2(\thetavec))\right]_{2,2}
	=\frac{\partial^2 \log\pi_2(\thetavec)}{\partial \theta_2^2}\hspace{1.75cm}
	\nonumber\\=
	- \frac{1}{(\pi_2(\thetavec))^2}\left(\frac{\partial \pi_2(\thetavec)}{\partial \theta_2}\right)^2
	+
	 \frac{1}{\pi_2(\thetavec)}\frac{\partial^2 \pi_2(\thetavec)}{\partial \theta_2^2}.
	\eeqna
	Similarly, 
	since $\pi_1(\thetavec)=1-\pi_2(\thetavec)$, 
	the only nonzero element of $\nabla_{\thetavecsmall_\Lambda}(\nabla^T_{\thetavecsmall_\Lambda}\log\pi_1(\thetavec))$
	is its $(2,2)$th element, which satisfies
		\beqna
	\label{der22_B}
	\left[\nabla_{\thetavecsmall_\Lambda}(\nabla^T_{\thetavecsmall_\Lambda}\log\pi_1(\thetavec))\right]_{2,2}
	=	\hspace{3.75cm}\nonumber\\
	- \frac{1}{(1-\pi_2(\thetavec))^2}\left(\frac{\partial \pi_2(\thetavec)}{\partial \theta_2}\right)^2
	-
	 \frac{1}{1-\pi_2(\thetavec)}\frac{\partial^2 \pi_2(\thetavec)}{\partial \theta_2^2}.
	\eeqna
The closed-form expressions of \eqref{der22} and \eqref{der22_B} are obtained by realizing that \cite{brychkov2012some}
	\be
	\label{derive}
	\frac{\partial}{\partial \theta_2}\pi_2(\thetavec)=
	\frac{\lambda}{\theta_2}\left(
	Q_{\frac{3}{2}}(\sqrt{\lambda},\sqrt{\gamma})
	-Q_{\frac{1}{2}}(\sqrt{\lambda},\sqrt{\gamma})
	\right)
	\ee
and
	\beqna
	\label{ccc}
	\frac{\partial^2}{\partial \theta_2^2}\pi_2(\thetavec)
		=-\frac{ \lambda}{\theta_2^2}\left(
	Q_{\frac{3}{2}}(\sqrt{\lambda},\sqrt{\gamma})
	-Q_{\frac{1}{2}}(\sqrt{\lambda},\sqrt{\gamma})
	\right)\nonumber\\
		+\frac{ \lambda^2}{\theta_2^2}\left(Q_{\frac{5}{2}}(\sqrt{\lambda},\sqrt{\gamma})-2 Q_{\frac{3}{2}}(\sqrt{\lambda},\sqrt{\gamma})+	Q_{\frac{1}{2}}(\sqrt{\lambda},\sqrt{\gamma})\right).
	\eeqna
	%https://www.tandfonline.com/doi/pdf/10.1080/10652469.2011.573184?needAccess=true

Based on the Gaussian pdf and \eqref{pi2_reg},  it can be shown, similarly to in Appendix \ref{sparseAPP}, that
the function $	f(\xvec |{\hat{\Lambda}=\Lambda}_k; \thetavec_\Lambda)$ is smooth in the sense that its first- and second-order derivatives are well defined. The associated second-order derivatives appear in 
\eqref{FIM_GLM}, \eqref{der22_B}, and \eqref{ccc}.
In addition, based on the considered model and \eqref{bayes}, the support of the conditional pdf, $f(\xvec|{\hat{\Lambda}=\Lambda}_k;\thetavec_{\Lambda}) $, is $\mathbb{R}^{N}$
 for all   $\thetavec_\Lambda$, which implies that Condition \ref{cond2} is satisfied.
Thus, 
	 Conditions \ref{cond1}-\ref{cond4} are satisfied  for the considered model as long as the selective FIMs are  nonsingular matrices, as required in Condition \ref{cond1}.
By substituting \eqref{60}-\eqref{der22_B} in \eqref{Jk_reg} and then substituting the result in
\eqref{BsCRB},
we obtain that  
	the selective CRB for this case is given by
	\beqna
	\label{BsCRB_sim}
		\Bmat_{\text{sCRB}}(\thetavec)
		=
\left[\begin{array}{cc}
\pi_1(\thetavec)[\Jmat_1^{-1}(\thetavec)]_{1,1}&0\\
0&0\end{array}\right]
+\pi_2(\thetavec)\Jmat_2^{-1}(\thetavec),
	\eeqna
where
	\beqna
	\label{Jk_reg2}
	\mathbf{J}_k(\thetavec)
	=\frac{1}{\sigma^2}{\Hmat_2}^T{\Hmat_2}+
	\left[\begin{array}{cc}
0&0\\
0&\frac{\partial^2 \log\pi_k(\thetavecsmall)}{\partial \theta_2^2}
\end{array}\right],~k=1,2.
	\eeqna
Finally,
by using the definition in \eqref{pm_def}, it can
	 be seen that
	$
	p_1(\thetavec)=1
	$
	and $p_2(\thetavec)=\pi_2(\thetavec)$.
	Thus,
 the selective CRB on the  trace MSE  of selective unbiased estimators from \eqref{sCRBtrace2} for this case is given by
		\beqna
		\label{sCRBtrace_gic}
			{\text{Tr}}\left({\text{MSE}}(\hat{\thetavec},\thetavec_{\Lambda})\right)	\geq\hspace{4.75cm}
						\nonumber\\
		\pi_1(\thetavec)
			[\Jmat_1^{-1}(\thetavec)]_{1,1}
			+\pi_2(\thetavec)\sum_{m=1}^2
			[\Jmat_2^{-1}(\thetavec)]_{m,m}
			+\pi_1(\thetavec)\theta_2^2,
		\eeqna
under the assumption that $\Jmat_1(\thetavec)$ and $\Jmat_2(\thetavec)$ are nonsingular matrices. This assumption held for all tested scenarios.

\begin{figure}[htb]
		\begin{center}
          \begin{tabular}[t]{c}
            %\subfigure[]{\includegraphics[width=8.25cm]{AIC}}
		%\\
		 \subfigure[]{\includegraphics[width=7.5cm]{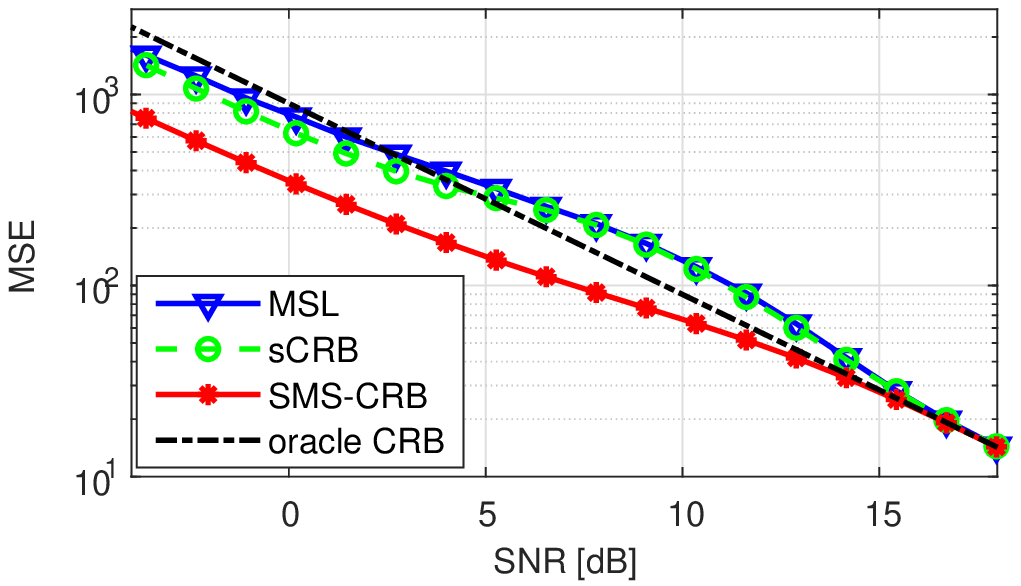}}
		\\
		\vspace{-0.5cm}
 \subfigure[]{\includegraphics[width=7.5cm]{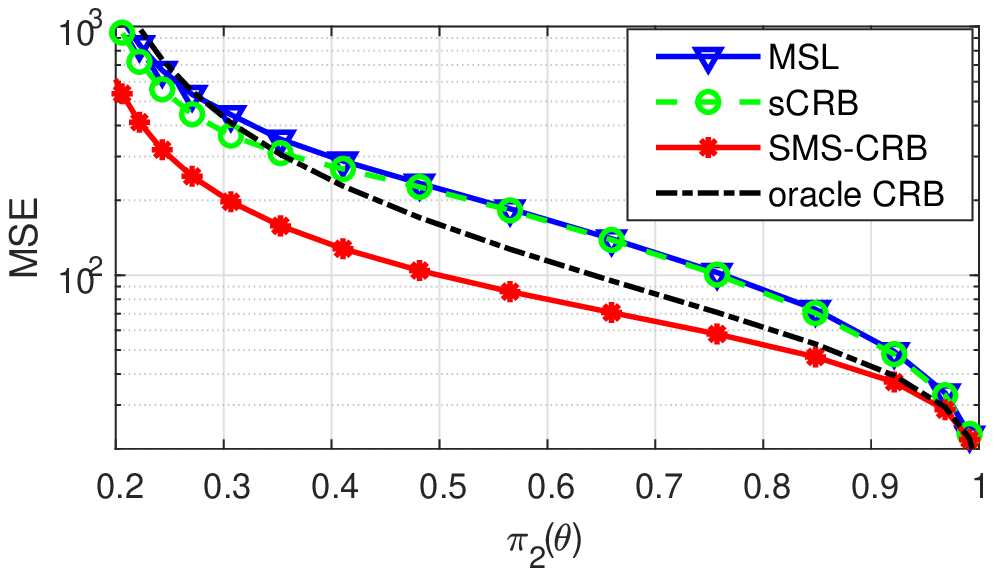}}
\end{tabular}
        \end{center}
		\caption{General linear model with AIC selection rule: the MSE of the MSL  estimator, the selective CRB,  the SMS-CRB, and the oracle CRB,
			 versus SNR  (a)  and 
			versus the probability of selection of the true model, $\pi_2(\thetavec)$ (b). }
		\label{AICfigs}	
	\end{figure}

	First, we show the results for the AIC model selection rule, i.e. where $\tau\left(N,|\Lambda_k|\right)=2$ in \eqref{GIC}.
	The  selective CRB from
	\eqref{sCRBtrace_gic},
 the trace of the SMS-CRB from  \cite{Sando_Mitra_Stoica2002}, and the trace of the oracle CRB for the $k=2$ model are evaluated and compared to the MSE of the MSL estimator,  where the selection of the likelihood is based on the AIC criterion, versus signal-to-noise ratio (SNR) and versus $\pi_2(\thetavec)$  in 
Figs. \ref{AICfigs}.a and  \ref{AICfigs}.b, respectively.
The SNR is defined as ${\rm{SNR}} \define 10\log 10 \frac{||\Hmat\thetavecsmall||^2}{N\sigma^2}$,
	where $N=1,500$ samples,  $\thetavec=[4,-3]^T$, $\hvec_1=[1,\ldots,1]^T$, and the values of $\hvec_2$ are 
	randomly drawn from a uniform distribution in the interval  $[0,10]$.
	It can be seen that the oracle CRB is not a valid bound on the MSE of the MSL estimator for low SNRs, in contrast to the selective CRB and the SMS-CRB.
	Moreover, the proposed selective CRB is tighter than the SMS-CRB
	and, in this example,
	can predict the ``breakdown phenomena", i.e. the threshold region 
	where the MSE of the MSL estimator deviates from the 
	 oracle CRB.
	 This breakdown means that the estimator makes gross, anomalous errors 
	 with a high probability \cite{Merhav_2011}, where in this case these errors are in the model selection phase. 

\begin{figure}[htb]
		\vspace{-0.25cm}
		\hspace{-0.75cm}
		\centering\includegraphics[width=8.5cm]{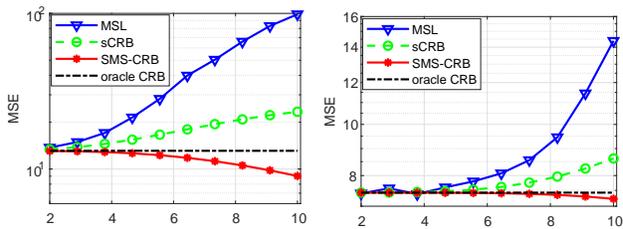}
		\vspace{-0.6cm}
		\caption{ General linear model with GIC selection rule: The MSE of the MSL  estimator,   the selective CRB,  and the SMS-CRB,
			versus different values of the parameter $\tau(N,|\Lambda_k|)$, with SNR$=-3.5$dB (left) and $0$dB (right).}
		\vspace{-0.5cm}
		\label{GICc}	
	\end{figure}

In Fig. \ref{GICc}  
 we examined these bounds and the performance of the MSL estimator for different values of
the GIC penalty term,
	$\tau(N,|\Lambda_k|)$, from \eqref{GIC},
ranging from  $\tau(N,|\Lambda_k|)=2$, associated with the AIC, to $\tau(N,|\Lambda_k|)=\log N$, associated with the MDL, for $N=150$ and SNR$=-3.5,0$dB.
The oracle CRB is a constant for any penalty, $\tau(N,|\Lambda_k|)$;
thus, it is inappropriate for post-model-selection estimation.
Moreover,
the SMS-CRB decreases as $\tau(N,|\Lambda_k|)$ increases, in contrast to the MSE of the MSL estimator. This is since the assumptions of the SMS-CRB (Assumptions \ref{A1}-\ref{A2}) do not hold in this case. 
 It can be seen that the selective CRB is tighter than the SMS-CRB and the oracle CRB in all examined scenarios,  but it is less tight
as  $\tau(N,|\Lambda_k|)$ increases.
The gap between the selective CRB and the MSE of the MSL estimator may suggest that
 better post-model-selection estimators can be derived, such as conditional ML estimators \cite{meir2017tractable,RoutTong2016,Harel_Routtenberg_2019}.
Since in this figure we present only the MSE of the correct parameters, the penalty on  overestimation is not presented in this figure. 
Since the AIC tends to overestimate the order of the model,
the MSL estimator of the  likelihood selected by the AIC rule
 has a lower MSE than the MSL estimator of the  likelihood selected by the MDL rule.
As a result, the selective CRB is tighter with the AIC selection rule than with the MDL selection rule.
This is in line with previous observations:
``the behavior of AIC with respect to the probability
of correct detection is not entirely satisfactory.
Interestingly, it is precisely this kind of behavior that
appears to make AIC perform satisfactorily with respect
to the other possible type of performance measure" \cite{stoica2004model}.
%%%%%%%%%%%%%%

	\subsection{Example 2 - Sparse vector estimation}
	\label{sparse_ex}
	In this subsection, we demonstrate the use of the selective CRB for measuring the achievable MSE in the sparse estimation problem from Section \ref{sparse}. 
	We validate in simulations the assumption that the matrices $\Amat_{\hat{\Lambda}}^T\Amat_{\hat{\Lambda}}$, and $\Amat_{\hat{\Lambda}}^T\Amat_{\hat{\Lambda}}+\Amat_\Lambda^T\Qmat_k\Amat_\Lambda$, $k=1,\ldots,K$, are nonsingular matrices.
The ML sparse estimator
is computationally prohibitive when the dimensions are large \cite{sparse_con}, and, thus, cannot be used in practice.
	The coherent MSL  estimator, which maximizes the likelihood selected by the OST selection rule for this model,
	is given by (see, e.g. \cite{sparse_con})
	\be
	\label{ML_reg_sparse}
	\hat{\thetavec}^{{\text{ML|OST}}}_{\hat{\Lambda}}
	=\left(\Amat_{\hat{\Lambda}}^T\Amat_{\hat{\Lambda}}\right)^{-1}\Amat_{\hat{\Lambda}}^T\xvec,
	\ee
 where $\hat{\Lambda}$ is the estimated support set by the OST rule, defined in \eqref{sparseSelection}. The other parameters of this estimator are set to zero, 
i.e. 		$\hat{\thetavec}^{{\text{ML|OST}}}_{\hat{\Lambda}^c}=\zerovec$.
Thus, according to 
\eqref{coherency_gen}, the MSL is a coherent estimator.
In order to have a fair comparison with the oracle CRB, the results in this subsection
%%%%%and  \ref{sparse_figs2} 
are the MSE of the true parameters, ${\rm{E}}_{\thetavecsmall_\Lambda}[(\hat{\thetavec}_{\Lambda}-\thetavec_{\Lambda})(\hat{\thetavec}_{\Lambda}-\thetavec_{\Lambda})^T]$. Similarly, the bounds in these figures are on the submatrix of the MSE of the true parameters.

\begin{figure}[htb]
		\begin{center}
{\includegraphics[width=7.75cm]{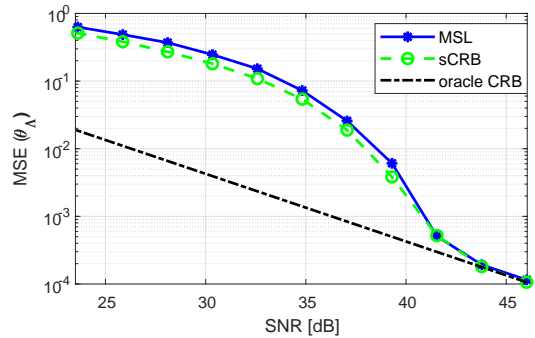}}
        \end{center}
		\caption{
		Sparse vector estimation:  The MSE of the MSL  estimator, where the likelihood is selected by  the OST rule, the selective CRB, and the oracle CRB,
			 versus SNR.   }
		\label{sparse_figs}	
	\end{figure}

First, 
	we generate a random $7\times 14$ dictionary, $\Amat$, from a zero-mean Gaussian i.i.d. distribution, whose columns $\avec_m$ were normalized so that $\avec_m^T\avec_m$=1, $m=1,\ldots,M$,
	where $M=14$ and the mutual coherence  of the chosen matrix  \cite{donoho2003optimally},
	which is the maximum absolute value of the normalized inner product between the columns  of the matrix,
	was $\mu=0.5673$.
	We choose a support set uniformly at random, set
$\thetavec_{\Lambda}={\mathbf{1}}$, and change the value of $\sigma^2$ to obtain different SNR values.
The OST selection rule \eqref{sparseSelection} is implemented with $c=0.95$ as a threshold.
The MSE  of the
 true parameters
of the
MSL estimator from \eqref{ML_reg_sparse} is compared with the selective CRB for sparse estimation, computed by using the selective FIM from Theorem \ref{Th2} in Fig.  \ref{sparse_figs} for a support set size of $|\Lambda| = 3$ and different noise variances, $\sigma^2$.
It can be seen  that
the MSE of the  MSL estimator and the proposed selective CRB converge asymptotically to the oracle CRB.  
Moreover, for this scenario, the selective CRB is a tight and valid bound on the performance of  the  MSL estimator   and
  it  predicts the ``breakdown phenomena".
  This is since it uses the accurate probability of selection and since in this case asymptotically, the OST chooses the correct model. 
In contrast, the oracle CRB is not an  informative bound in the non-asymptotic region and does not predict the threshold.
The SMS-CRB cannot be calculated for this case since it requires that the FIM will be a nonsingular matrix for any model, which is not the case in  general sparse estimation.
A discussion on the insights behind the behavior of the proposed selective CRB 
and the influence of the selection probability appears in Subsection \ref{digging}. 
	
%%The MSE of the MSL and the CRBs were also compared for varying sparsity levels (support set sizes). 
		%%%%%In order to also compare  with the SMS-CRB, in this experiment we set the  dictionary, $\Amat$, to be the
		 %%%%%$16\times 16$ Hadamard matrix with normalized columns, which is a full-rank matrix.
		%%%%%Then, the simulation was repeated for different support set sizes in the range $1\leq |\Lambda|\leq 16$, with a constant noise standard deviation of $\sigma=0.1594$. 
		%%%%%The results are plotted in Fig.   \ref{sparse_figs2}.
  %%%%%It can be seen that the SMS-CRB and the oracle CRB are valid bounds but are not tight.
	%%%%%In contrast, the proposed bound is almost identical to the MSE of the MSL estimator  with the likelihood selected by the OST rule.
	%%%%%As the  size of the support set increases,  more parameters contribute to the MSE, and, thus, the MSE increases.
	%%%%%In addition, as the  size of the support set increases,
	%%%%%correct support set recovery
%%%%%becomes more difficult, which increases the sensitivity of the estimate to random
%%%%%measurement fluctuations. 
	%%%%%\begin{figure}[htb]
		%%%%%\begin{center}
%%%%%{\includegraphics[width=7.5cm]{hadamard17_2.eps}}
        %%%%%\end{center}
				%%%%%\vspace{-0.25cm}
		%%%%%\caption{
		%%%%%Sparse vector estimation:  The MSE of the MSL  estimator where the likelihood is selected by  the OST rule, the selective CRB,
		%%%%%the SMS-CRB, and the oracle CRB,
			 %%%%%versus $|\Lambda|$}.  
		%%%%%\label{sparse_figs2}	
	%%%%%\end{figure}
	
Since for general sparse representation problems the MSL estimator is not a selective-unbiased estimator, the proposed selective CRB may not be tight for the general case and, 
	in some cases, may be even higher than the actual MSE.
Thus,  in the following, we demonstrate the use of the  biased selective CRB  from \eqref{Rdef} for this case, with $\Amat=\Imat$ and $L=M$. 
	 For this case,
by substituting \eqref{ML_reg_sparse} and \eqref{sparseSelection} in \eqref{11_bias} and using the independency of  the elements of the noise vector, $\wvec$, from \eqref{sparseMeas}, we obtain that the $m$th element of the selective bias
of the MSL  estimator, 
 under the selected $k$th model,  is
		\beqna
		\label{try17}
		[\bvec_k(\thetavec,\Lambda)]_m&=&
		{\rm{E}}_{\thetavecsmall_\Lambda}\left[\left.x_m-\theta_m\right||x_m|>c\right]
		\nonumber\\
&=&-\sigma\frac{\phi(\beta_m)
-\phi(\alpha_m)}{1-\Phi(\alpha_m)+\Phi(\beta_m)},
\eeqna
$\forall k=1,\ldots,K$, such that $m\in \Lambda_k$,
		where, according to \eqref{alpha_def} and \eqref{bbb}, in this case 
	$
\alpha_m=\frac{c-\theta_m}{\sigma}$ and $\beta_m=\frac{-c-\theta_m}{\sigma}$, $\forall m=1,\ldots,M$.
The last equality in \eqref{try17} is obtained 
				by using known results on the moments of  truncated Gaussian distributions  \cite{johnson1995continuous}, as well as \eqref{foldedCDF} from Appendix \ref{sparseAPP}.
				In addition, $[\bvec_k(\thetavec,\Lambda)]_m=0$, $\forall k=1,\ldots,K$, such that $m\in \Lambda_k^c$. 
		By substituting \eqref{try17} in \eqref{der_bias}, we obtain that the elements of the bias gradient matrix are
			\beqna
	\label{der_bias_sparse}
[\Gmat_k(\thetavec,\Lambda)]_{m,l}
=
-\sigma\frac{\partial}{\partial \theta_m} \frac{\phi(\beta_m)
-\phi(\alpha_m)}{1-\Phi(\alpha_m)+\Phi(\beta_m)} 
\nonumber\\
=
-\frac{\beta_m\phi(\beta_m)
-\alpha_m\phi(\alpha_m)}{1-\Phi(\alpha_m)+\Phi(\beta_m)} -\frac{1}{\sigma^2} ([\bvec_k(\thetavec,\Lambda)]_m)^2,
\eeqna
for any $m\in \Lambda_k$, such that $m=[\Lambda]_l$, and zero otherwise. 
	By substituting \eqref{spsJ} and \eqref{der_bias_sparse} in \eqref{Rdef}, we obtain the biased CRB for this 
	case. 
	We also evaluate   the performance of the 
	iterative post-selection conditional  ML (CML) from \cite{meir2017tractable}  for this case.
	%%%%%\beqna
	%%%%%\label{CML}
	%%%%%\hat{\thetavec}^{{\text{CML}}}=
	%%%%%\arg\max_{\thetavecsmall}\left\{
	%%%%%\frac{f(\xvec;\thetavec)}{\prod_{m\in \hat{\Lambda}} 
	%%%%%\Pr(|x_m|>c;\thetavec) }\right\}
	%%%%%=\arg\max_{\thetavecsmall}\nonumber\\
	%%%%%\left\{
%%%%%-\frac{1}{2\sigma^2}||\xvec-\thetavec||_2^2
%%%%%-\sum_{m\in \hat{\Lambda}}\log(1-\Phi(\alpha_m)+\Phi(\beta_m))\right\},
	%%%%%\eeqna
	%%%%%where the second equality is obtained by using the monotonicity of 
 %%%%%the likelihood function, substituting the model from Section \ref{sparse} with $\Amat=\Imat$, and  removing the constant terms that are not function of $\thetavec$.
%%%%The solution of \eqref{CML}
 Here we use a single iteration and initialize the CML estimator  by $\hat{\thetavec}^{{\text{ML|OST}}}_{\hat{\Lambda}}$.
The biased selective CRB for the CML estimator was evaluated numerically, by calculating the selective bias and its gradient, as defined in \eqref{11_bias} and \eqref{der_bias}.

%%%%%%numerical bound for cml

%%%%%\beqna
%%%%%		[\bvec_k(\thetavec,\Lambda)]_m=		{\rm{E}}_{\thetavecsmall_\Lambda}\left[\left.\hat{\theta}_m-\theta_m\right||x_m|>c\right]
%%%%%\eeqna
%%%%%\beqna
%%%%%	\Gmat_{m,m}=\frac{\partial}{\partial \theta_m}
%%%%%		{\rm{E}}_{\thetavecsmall_\Lambda}\left[\left.\hat{\theta}_m-\theta_m\right||x_m|>c\right]
%%%%%		\nonumber\\
%%%%%		=\frac{\partial}{\partial \theta_m}
%%%%%		\int_{x_m:|x_m|>c} \frac{f(x_m;\theta_m)}{\Pr(|x_m|>c;\theta_m)} (\hat{\theta}_m-\theta_m)\ud x_m
%%%%%			\nonumber\\
%%%%%		=\frac{\partial}{\partial \theta_m} \frac{1}{\Pr(|x_m|>c;\theta_m)}
%%%%%		\int_{x_m:|x_m|>c}  f(x_m;\theta_m)(\hat{\theta}_m-\theta_m)\ud x_m
%%%%%			\nonumber\\
%%%%%		=- \frac{\phi(\alpha_m)-\phi(\beta_m)}{\sigma \Pr^2(|x_m|>c;\theta_m)}
%%%%%		\int_{x_m:|x_m|>c}  f(x_m;\theta_m)(\hat{\theta}_m-\theta_m)\ud x_m
%%%%%			\nonumber\\
%%%%%	+ \frac{1}{\Pr(|x_m|>c;\theta_m)}
%%%%%	\nonumber\\
%%%%%	\times
%%%%%		\int_{x_m:|x_m|>c} \frac{1}{\sigma^2}(\hat{\theta}_m-\theta_m)(x-\theta_m) f(x_m;\theta_m)\ud x_m-1
%%%%%		\nonumber\\
%%%%%		=- \frac{\phi(\alpha_m)-\phi(\beta_m)}{\sigma(1-\Phi(\alpha_m)+\Phi(\beta_m))}
%%%%%		b_m
%%%%%			\nonumber\\
%%%%%	+ {\rm{E}}\left[\left. \frac{1}{\sigma^2}(\hat{\theta}_m-\theta_m)(x-\theta_m)
%%%%%	\right||x_m|>c\right]-1
%%%%%\eeqna

\begin{figure}[htb]
		\begin{center}
{\includegraphics[width=9cm]{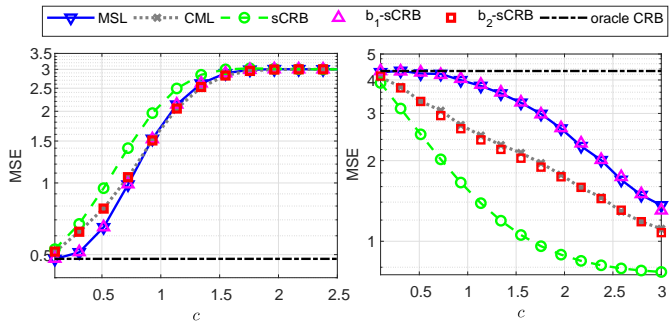}}
        \end{center}
				\vspace{-0.25cm}
		\caption{
		Sparse vector estimation:  The MSE of the MSL and of the CML estimators, where the likelihood is selected by  the OST rule, the selective CRB,
		the biased selective CRB with the MSL bias, ${\text{b}}_1$-sCRB, 
		the biased selective CRB with the CML bias, ${\text{b}}_2$-sCRB, and the oracle CRB,
			 versus the threshold, $c$, for
			1) $\theta_m=1, \sigma=0.4$ (left);
			2) $\theta_m=0.5, \sigma=1.2$ (right).}
		\label{sparse_figs3}	
	\end{figure}

	The MSE of the MSL and the CML, as well as  the selective CRB, the biased selective CRBs, and the SMS-CRB,  are shown in Fig.   \ref{sparse_figs3} for varying threshold values, $c$,
	$K=3$,
	$M=L=8$, and for two scenarios:
1) $\thetavec_{\Lambda}=1,~\sigma=0.4$;
and 2) $\thetavec_{\Lambda}=0.5,~\sigma=1.2$.
We denote the biased selective CRB associated with the bias of the MSL estimator by 
${\text{b}}_1$-sCRB  and the biased selective CRB associated with the bias of the CML estimator by 
${\text{b}}_2$-sCRB.
  It can be seen that for Scenario 1, the selective CRB is higher than the MSE of the MSL,
	while for Scenario 2
	it is lower than the MSE of the MSL, but is not tight.
	The performance of the biased selective CRB, ${\text{b}}_1$-sCRB, coincides with  the performance of the MSL estimator.
Similarly, the performance of the  biased selective CRB, ${\text{b}}_2$-sCRB, almost coincides with  the performance of the CML estimator. 
	In contrast to the oracle CRB, which does not show the influence of the threshold parameter, $c$, on the performance,
	the selective CRB bounds are informative and demonstrate similar patterns (w.r.t. $c$) to the MSE of the estimators. 
	It can be seen that the MSE of the CML estimator is closer to the (unbiased) selective CRB,
	which may be explained by the results from  \cite{meir2017tractable} that show that the CML has a lower selective bias compared with those of  the MSL estimator. 
	The SMS-CRB is not shown here since it is significantly lower than the other bounds
	and since it is also not a valid bound in many tested cases.
	%%%%%%%%%%%%%%%%%%%
\subsection{Discussion}
\label{digging}
It is well known that the threshold effect in the MSE of estimators in nonlinear models  is due to large error contributions to the total MSE \cite{athley2005threshold,BARANKIN,Merhav_2011}.
In the context of post-model-selection estimation, large errors can be interpreted as the errors due to the incorrect model selection, including both true parameters that are estimated as zero parameters and
the wrong selection of null parameters. 
The small/local errors can be interpreted as the errors due to the estimation approach, when the model selection approaches the true selection.
For example, 
if we look at the selective CRB in \eqref{sCRBtrace}, 
the influence of the large errors due to the preliminary selection approach 
 is expressed through the following aspects:
1)
the term $\sum_{m\in\Lambda}\left(1-p_m(\thetavec_{\Lambda})\right)\theta_m^2$, which is a large-error term, associated with true parameters that are missed by the selection rule; 2)
the bias term, $\bvec_k(\thetavec,\Lambda)$,  which represents large errors for the elements associated with the null parameters; and
3) the selective FIM, $\Jmat_k(\thetavec_{\Lambda})$, which, as shown in Lemma \ref{lemma1}, includes the influence of the second-order derivatives of the probability  of selecting the $k$th model,
$\pi_{k}(\thetavec_\Lambda)$, which is a large-error measure that is sensitive to the  selection stage. 

Different system parameters affect the ability of the proposed selective CRB to predict the threshold in estimation after model selection problems, which is due to
 the model selection stage.
For example,
in  Fig. \ref{GICc} it can be seen that the tightness is different for different selection rules.
Another example is for the sparse vector estimation from Subsection 
\ref{sparse_ex};
our results (not shown here) indicate that the mutual coherence of $\Amat$ affects the 
tightness of the proposed bound, where the selective CRB is less tight as the mutual coherence increases. This is reasonable, since  the mutual coherence is 
used as a measure of the ability of suboptimal algorithms  to correctly identify the true representation of a sparse signal. 
Intuitively,  In the extreme case when two columns are completely coherent, it will be impossible to distinguish between their contributions and, thus, impossible to recover the sparse signal, $\thetavec$.
%%%
In contrast to the selective CRB,
 the oracle CRB cannot predict the threshold which is due to the selection, since it does not take into account the coherency property of estimators, nor the preliminary selection stage and the uncertainty in the model. Similarly, the SMS-CRB from \eqref{SMS-CRB} cannot predict the threshold since it is developed under Assumption \ref{A1} that there is no underestimation of
the order, i.e. the probability that a true parameter has not been selected goes to zero.
	%%%%%%%%%%%%%%%%%%%%%%%%%%%%%%%%%%%%%%%%%%%%%%%%%%%%%%%%%%%%%%%%%%%%%%%%%%%%%%%%%%%%%%%%%%%%%%%%%%%%%%%%%%%%%%%%%
	\vspace{-0.25cm}
	\section{Conclusion}
	\label{conclusion}
	In this paper, we consider non-Bayesian parameter estimation after model selection. The selection of a model is interpreted here as estimating the selected parameters under a zero constraint on the  unselected parameters.
	First, the notation of  selective unbiasedness is defined by using  Lehmann's concept of unbiasedness. We propose a novel Cram$\acute{\text{e}}$r-Rao-type bound, denoted by selective CRB, which is a lower bound on the MSSE	of any coherent and selective unbiased estimator. 
		The selective CRB is used also to obtain a lower bound on  the MSE of coherent estimators.
	The relation between  the proposed selective CRB 
	and existing bounds, the SMS-CRB  and the misspecified CRB, are investigated. 
The selective CRB was derived for the special case of sparse vector estimation, where the  recovery of the support set is performed based on the OST selection rule.
	In simulations, we have demonstrated that the proposed selective CRB provides an informative bound on the
	MSL estimator, while the oracle CRB is not a valid bound for this case. Moreover, the proposed selective CRB is  tighter than the  SMS-CRB (when it exists), and, in some cases, it predicts the breakdown phenomena of the MSL estimator.  
The predictive capability  is attributed to the fact that the selective CRB takes into account the influence of large errors  caused by incorrect model selection and the coherency constraint.
	Our experimental results show that the tightness of the proposed selective CRB depends on the selection rule, on the threshold  of the selection rule, and on the measurement matrix.
%	For the sparse setting, the biased selective CRB has been shown to be tight bound on the MSE of the MSL estimator. 
	
	Topics for future research include 
	the derivation of  post-model-selection estimators in the sense of reduced bias, lower MSE, and achievability of the proposed selective CRB, as well as through analysis of the performance (bias and MSE) of the estimators of the null parameters.
	In addition, the selective CRB may be extended for a Bayesian selection rule
 and 
	 misspecified  models \cite{richmondHorowitz,vuong,pajovic,Fortunati_Gini_Greco2016,white1982maximum}.
	The development of
	 MSE approximations by the method of interval errors  \cite{VAN-TREES,Richmond2006,richmond2012aspects} 
	 could shed more light on the
MSE regions
of performance and on the threshold phenomenon. 
	Finally,  low-complexity bound approximations, such as an empirical CRB, should be developed for cases with intractable probability of selection.  
	%%%%%%%%%%%%%%%%%%%%%%%%%%%%%%%%%%%%%%%%%%%%%%%%%%%%%%%%%%%%%%%%%%%%%%%%%%%%%%%%%%%%%%%%%%%%%%%%%%%%
	\section{ACKNOWLEDGEMENT}
The authors would like to thank the anonymous reviewers for
their valuable comments.
{\textcolor{black}{In particular, based on the review process, we extended the SSE cost function from   \cite{meirRoutSSP} to also include also the influence of the null parameters.}}
%%%%%%%%%%%%%
	\appendices
	\renewcommand{\thesectiondis}[2]{\Alph{section}:}
	\section{Proof of Proposition \ref{propLehm}}
	\label{unbiasApp}
In this appendix, we prove that the selective unbiasedness is obtained from the Lehmann unbiasedness with the SSE cost function.
By substituting  (\ref{aSSE}) in (\ref{Lehmann_vector}), 
we obtain that the  Lehmann unbiasedness for the SSE cost function requires that
	\beqna
	\label{kMin_org}
		{\rm{E}}_{\thetavecsmall_\Lambda}\left[(\hat{\thetavec}_{\hat{\Lambda}}\zp-\etavec_{\hat{\Lambda}}\zp)(\hat{\thetavec}_{\hat{\Lambda}}\zp-\etavec_{\hat{\Lambda}}\zp)^T\right]\hspace{2cm}\nonumber\\
		\succeq 	
	{\rm{E}}_{\thetavecsmall_\Lambda}\left[(\hat{\thetavec}_{\hat{\Lambda}}\zp-\thetavec_{\hat{\Lambda}}\zp)(\hat{\thetavec}_{\hat{\Lambda}}\zp-\thetavec_{\hat{\Lambda}}\zp)^T\right],
	\eeqna
	$\forall \etavec,\thetavec\in {\mathbb{R}}^M$, where $\thetavec_{\Lambda^c}=\zerovec$.
	By using  \eqref{bayes} and the law of total expectation, \eqref{kMin_org} can be rewritten as
		\beqna
	\label{kMin}
		\sum\limits_{k=1}^K\pi_{k}(\thetavec_{\Lambda})
		{\rm{E}}_{\thetavecsmall_\Lambda}\left[(\hat{\thetavec}_{\Lambda_k}\zp-\etavec_{\Lambda_k}\zp)(\hat{\thetavec}_{\Lambda_k}\zp-\etavec_{\Lambda_k}\zp)^T|{\hat{\Lambda}=\Lambda}_k\right]\nonumber\\
		\succeq \sum\limits_{k=1}^K\pi_{k}(\thetavec_{\Lambda})	
	{\rm{E}}_{\thetavecsmall_\Lambda}\left[(\hat{\thetavec}_{\Lambda_k}\zp-\thetavec_{\Lambda_k}\zp)(\hat{\thetavec}_{\Lambda_k}\zp-\thetavec_{\Lambda_k}\zp)^T|{\hat{\Lambda}=\Lambda}_k\right],
	\eeqna
By adding and subtracting
 \be
 \label{uuu}
 \uvec_k(\thetavec_{\Lambda})\define {\rm{E}}_{\thetavecsmall_\Lambda}[\hat{\thetavec}_{\hat{\Lambda}}\zp|{\hat{\Lambda}=\Lambda}_k]
 ={\rm{E}}_{\thetavecsmall_\Lambda}[\hat{\thetavec}_{{\Lambda}_k}\zp|{\hat{\Lambda}=\Lambda}_k]
 \ee
 inside the
  two brackets associated with the $k$th model on both sides of
 \eqref{kMin} for any $k=1,\ldots,K$,
 we obtain
 \beqna
	\label{kMin_with_k}
		\sum\limits_{k=1}^K\pi_{k}(\thetavec_{\Lambda})
		{\rm{E}}_{\thetavecsmall_\Lambda}\left[\left(\hat{\thetavec}_{{\Lambda}_k}\zp
		-\uvec_k(\thetavec_{\Lambda})
		+\uvec_k(\thetavec_{\Lambda})
		-\etavec_{{\Lambda}_k}\zp\right)\right.
		\nonumber\\
			\left.
		\times \left(\hat{\thetavec}_{{\Lambda}_k}\zp-\uvec_k(\thetavec_{\Lambda})+\uvec_k(\thetavec_{\Lambda})-\etavec_{{\Lambda}_k}\zp\right)^T|{\hat{\Lambda}=\Lambda}_k\right]\nonumber\\
		\succeq \sum\limits_{k=1}^K\pi_{k}(\thetavec_{\Lambda})	
	{\rm{E}}_{\thetavecsmall_\Lambda}\left[\left(\hat{\thetavec}_{{\Lambda}_k}\zp-\uvec_k(\thetavec_{\Lambda})
	+\uvec_k(\thetavec_{\Lambda})-\thetavec_{{\Lambda}_k}\zp\right)\right.
	\nonumber\\
	\left.
	\times\left(\hat{\thetavec}_{{\Lambda}_k}\zp-\uvec_k(\thetavec_{\Lambda})+\uvec_k(\thetavec_{\Lambda})-\thetavec_{{\Lambda}_k}\zp\right)^T|{\hat{\Lambda}=\Lambda}_k\right],
	\eeqna
	$\forall \etavec,\thetavec\in {\mathbb{R}}^M$.
 First, we note that we can remove the identical terms,
 \be
 \label{termK}
 \sum\limits_{k=1}^K\pi_{k}(\thetavec_{\Lambda}){\rm{E}}_{\thetavecsmall_\Lambda}\left[(\hat{\thetavec}_{{\Lambda}_k}\zp
		-\uvec_k(\thetavec_{\Lambda}))
		(\hat{\thetavec}_{{\Lambda}_k}\zp
			-\uvec_k(\thetavec_{\Lambda}) 
	)^T|{\hat{\Lambda}=\Lambda}_k\right],
\ee
  from the two sides of \eqref{kMin_with_k}.
  In addition,  by using 
   \eqref{uuu},
   it can be verified that 
  \beqna
  \label{zero_cross}
  	{\rm{E}}_{\thetavecsmall_\Lambda}\left[(\hat{\thetavec}_{{\Lambda}_k}\zp
		-\uvec_k(\thetavec_{\Lambda})) (\uvec_k(\thetavec_{\Lambda})-\etavec_{{\Lambda}_k}\zp)^T|{\hat{\Lambda}=\Lambda}_k\right]=\zerovec,
  \eeqna
  $\forall k=1,\ldots,K$.
  Thus, by subtracting \eqref{termK}  from the two sides of \eqref{kMin_with_k} and substituting \eqref{zero_cross}, we obtain that
  \eqref{kMin_with_k} can be rewritten as
   \beqna
	\label{kMin_with_k2}
		\sum\limits_{k=1}^K\pi_{k}(\thetavec_{\Lambda})
		\left(\uvec_k(\thetavec_{\Lambda})
		-\etavec_{{\Lambda}_k}\zp\right) \left(\uvec_k(\thetavec_{\Lambda})-\etavec_{{\Lambda}_k}\zp\right)^T\hspace{0.75cm}\nonumber\\
		\succeq \sum\limits_{k=1}^K\pi_{k}\left(\uvec_k(\thetavec_{\Lambda})-\thetavec_{{\Lambda}_k}\zp\right)
	\left(\uvec_k(\thetavec_{\Lambda})-\thetavec_{{\Lambda}_k}\zp\right)^T,
	\eeqna
	$\forall \etavec,\thetavec\in {\mathbb{R}}^M$.

Now, we note that by using 
 \eqref{uuu}, we obtain that the term on the r.h.s. of \eqref{kMin_with_k2}
 \beqna
 \label{bbbb}
\uvec_k(\thetavec_{\Lambda})-\thetavec_{{\Lambda}_k}\zp
={\rm{E}}_{\thetavecsmall_\Lambda}[\hat{\thetavec}_{\hat{\Lambda}}\zp-\thetavec_{\hat{\Lambda}}\zp|{\hat{\Lambda}=\Lambda}_k]
=\bvec_k(\thetavec,\Lambda),
\eeqna
where the last equality is obtained by substituting \eqref{11_bias}.
	By substituting \eqref{bbbb} in \eqref{kMin_with_k2}, we obtain that the unbiasedness condition can be written as 
	 \beqna
	\label{kMin_with_k3}
		\sum\limits_{k=1}^K\pi_{k}(\thetavec_{\Lambda})
		(\uvec_k(\thetavec_{\Lambda})
		-\etavec_{{\Lambda}_k}\zp) (\uvec_k(\thetavec_{\Lambda})-\etavec_{{\Lambda}_k}\zp)^T\hspace{1cm}\nonumber\\
		\succeq \sum\limits_{k=1}^K\pi_{k}\bvec_k(\thetavec,\Lambda)
	\bvec_k^T(\thetavec,\Lambda),
	\eeqna
	$\forall \etavec,\thetavec\in {\mathbb{R}}^M$.
Due to the non-negativity of the probabilities,
$\pi_{k}(\thetavec_{\Lambda})$, $k=1,\ldots,K$, and of the matrices on the l.h.s. of \eqref{kMin_with_k3} (in the sense of positive
semidefiniteness),	
	the l.h.s. of \eqref{kMin_with_k3} is necessarily a positive semidefinite  matrix. 
Therefore, 	
a sufficient condition for \eqref{kMin_with_k2} to hold is if \eqref{11_bias_zero} holds.

It should be noted that if  the Lehmann condition in \eqref{kMin}
	  must only be satisfied for  $\etavec$ with the same support set as $\thetavec$, i.e. for $\etavec$ with $\etavec_{\Lambda^c}=\zerovec$, then we will get a less restrictive selective unbiasedness condition.
	  In particular, in Proposition 1 in \cite{meirRoutSSP} 
	 the selective unbiasedness only restricts the values that are in the intersection of the true and estimated support, i.e. it only restricts the values of $\hat{\thetavec}_{\Lambda\cap\hat{\Lambda}}$.

		%%%%%%%%%%%%%%%%
		\section{Proof of Proposition \ref{prop_relation}}
		\label{MSE_der_app}
		By using Definition \ref{ZPdef}, the estimation error, $\hat{\thetavec}-\thetavec$, can be decomposed w.r.t. the selected support set,
	$\hat{\Lambda}$, and its complementary, $\hat{\Lambda}^c$, as follows:
	\be
	\label{ZPdemo}
		\hat{\thetavec}-\thetavec=\hat{\thetavec}_{\hat{\Lambda}}\zp-\thetavec_{\hat{\Lambda}}\zp+\hat{\thetavec}_{\hat{\Lambda}^c}\zp-\thetavec_{\hat{\Lambda}^c}\zp,
	\ee
	where all the vectors in  (\ref{ZPdemo}) have the same dimension, $M$. 
	By substituting \eqref{MSSE} and \eqref{ZPdemo} in the MSE matrix from \eqref{MSE_def}, one obtains
	\beqna
	\label{SSE}
		{\text{MSE}}(\hat{\thetavec},\thetavec,\Lambda)= {\rm{E}}_{\thetavecsmall_\Lambda}\left[\Cmat(\hat{\thetavec},\hat{\Lambda},\thetavec_{\Lambda})\right]\hspace{2cm} \nonumber\\
		+ {\rm{E}}_{\thetavecsmall_\Lambda}\left[\left(\hat{\thetavec}_{\hat{\Lambda}^c}\zp-\thetavec_{\hat{\Lambda}^c}\zp\right)\left(\hat{\thetavec}_{\hat{\Lambda}}\zp-\thetavec_{\hat{\Lambda}}\zp\right)^T\right]
		\hspace{0.35cm}\nonumber\\
		+{\rm{E}}_{\thetavecsmall_\Lambda}\left[\left(\hat{\thetavec}_{\hat{\Lambda}}\zp-\thetavec_{\hat{\Lambda}}\zp\right)\left(\hat{\thetavec}_{\hat{\Lambda}^c}\zp-\thetavec_{\hat{\Lambda}^c}\zp\right)^T\right]	\hspace{0.35cm}\nonumber\\
		+{\rm{E}}_{\thetavecsmall_\Lambda}\left[\left(\hat{\thetavec}_{\hat{\Lambda}^c}\zp-\thetavec_{\hat{\Lambda}^c}\zp\right)\left(\hat{\thetavec}_{\hat{\Lambda}^c}\zp-\thetavec_{\hat{\Lambda}^c}\zp\right)^T\right].
	\eeqna
	The coherency property from \eqref{coherency_gen} after a zero-padding approach, implies that $	\hat{\thetavec}_{\hat{\Lambda}^c}\zp=\zerovec$.
	By substituting $	\hat{\thetavec}_{\hat{\Lambda}^c}\zp=\zerovec$ in \eqref{SSE}, we obtain that for any coherent estimator, the MSE satisfies
		\beqna
	\label{SSE2}
		{\text{MSE}}(\hat{\thetavec},\thetavec_{\Lambda})= {\rm{E}}_{\thetavecsmall_\Lambda}\left[\Cmat(\hat{\thetavec},\hat{\Lambda},\thetavec_{\Lambda})\right]\hspace{2.75cm}\nonumber \\
		-{\rm{E}}_{\thetavecsmall_\Lambda}\left[\thetavec_{\hat{\Lambda}^c}\zp\left(\hat{\thetavec}_{\hat{\Lambda}}\zp-\thetavec_{\hat{\Lambda}}\zp\right)^T\right]
	-{\rm{E}}_{\thetavecsmall_\Lambda}\left[\left(\hat{\thetavec}_{\hat{\Lambda}}\zp-\thetavec_{\hat{\Lambda}}\zp\right)\left(\thetavec_{\hat{\Lambda}^c}\zp\right)^T\right]\nonumber\\
		+{\rm{E}}_{\thetavecsmall_\Lambda}\left[\thetavec_{\hat{\Lambda}^c}\zp\left(\thetavec_{\hat{\Lambda}^c}\zp\right)^T\right].\hspace{
	5cm}
	\eeqna
	Now, by using the law of total expectation, it can be seen that for any   coherent and  selective biased estimator, 
	the second term on the r.h.s. of \eqref{SSE2} satisfies
	\beqna
		\label{mixed_term}
			{\rm{E}}_{\thetavecsmall_\Lambda}\left[{\thetavec}_{\hat{\Lambda}^c}\zp\left(\hat{\thetavec}_{\hat{\Lambda}}\zp-\thetavec_{\hat{\Lambda}}\zp\right)^T\right]\nonumber\hspace{3cm}\\
		%=\sum_{k=1}^K\pi_{k}(\thetavec_\Lambda)
	%	{\rm{E}}_{\thetavecsmall_\Lambda}\left[{\thetavec}_{\Lambda_k^c}\zp(\hat{\thetavec}_{{\Lambda}_k}\zp-\thetavec_{{\Lambda}_k}\zp)^T|{\hat{\Lambda}=\Lambda}_k\right]
	%	\nonumber\\
		=\sum_{k=1}^K\pi_{k}(\thetavec_\Lambda)
		{\thetavec}_{\Lambda_k^c}\zp{\rm{E}}_{\thetavecsmall_\Lambda}\left[(\hat{\thetavec}_{{\Lambda}_k}\zp-\thetavec_{{\Lambda}_k}\zp)^T|{\hat{\Lambda}=\Lambda}_k\right]
		\nonumber\\
		=\sum_{k=1}^K\pi_{k}(\thetavec_\Lambda)
		{\thetavec}_{\Lambda_k^c}\zp	{\bvec_k^T(\thetavec,\Lambda}),\hspace{2.4cm}
		\eeqna
		where the last equality is obtained by substituting the selection bias property from (\ref{11_bias}).
		Similarly, by using the law of total expectation, the last term of the r.h.s. of \eqref{SSE2}
		satisfies
		\beqna
		\label{last_term}
	{\rm{E}}_{\thetavecsmall_\Lambda}\left[{\thetavec}_{\hat{\Lambda}^c}\zp({\thetavec}_{\hat{\Lambda}^c}\zp)^T\right]
		=\sum_{k=1}^K\pi_{k}(\thetavec_\Lambda){\thetavec}_{\Lambda_k^c}\zp({\thetavec}_{\Lambda_k^c}\zp)^T.
\eeqna
		By substituting \eqref{mixed_term} and \eqref{last_term} in \eqref{SSE2}
	we obtain that the MSE of a coherent and selective unbiased estimator is given by \eqref{MSE}.

%%%%%%%%%%%%%		
	\section{Proof of Theorem \ref{Th1}}
	\label{CRBProof}
	By using Conditions \ref{cond1} and \ref{cond2},
	it can be shown, similarly to in the proof of  Lemma 2.5.3 in \cite{point_est} for the unconditional likelihood function, that
\beqna
\label{zero_val}
{\rm{E}}_{\thetavecsmall_{\Lambda}}\left[\upsilonvec_k^T(\xvec,\thetavec_{\Lambda})|{\hat{\Lambda}=\Lambda}_k\right]\hspace{3cm}\nonumber\\
={\rm{E}}_{\thetavecsmall_{\Lambda}}\left[\nabla_{\thetavecsmall_{\Lambda}}\log f(\xvec|{\hat{\Lambda}=\Lambda}_k;\thetavec_{\Lambda})|{\hat{\Lambda}=\Lambda}_k\right]=
\zerovec,
\eeqna
$\forall k=1,\ldots,K$ and for any $\thetavec_{\Lambda}\in\mathbb{R}^{|\Lambda|}$,
where $\upsilonvec_k(\xvec,\thetavec_{\Lambda})$ is defined in \eqref{l_define}.
In addition, for any square-integrable function,  $\xsivec:\Omega_{\xvec}\times\Omega_{\thetavecsmall}\rightarrow{\mathbb{R}^{M}}$, the covariance inequality (see, e.g.   p. 33 in \cite{VAN-TREES}) implies that
\beqna \label{eq:vec_covariance_inequality}
{\rm{E}}_{\thetavecsmall_{\Lambda}}\left[\xsivec(\xvec,\thetavec)\xsivec^{T}(\xvec,\thetavec)\right]\succeq
{\rm{E}}_{\thetavecsmall_{\Lambda}}\left[\xsivec(\xvec,\thetavec)\upsilonvec_k^T(\xvec,\thetavec_{\Lambda})\right]\hspace{0.4cm}\nonumber\\
\times \Jmat_k^{-1}(\thetavec_{\Lambda}){\rm{E}}_{\thetavecsmall_{\Lambda}}\left[\upsilonvec_k(\xvec,\thetavec_{\Lambda})\xsivec^{T}(\xvec,\thetavec)\right],
\eeqna
where, according to Condition \ref{cond1}, the selective FIMs, $\Jmat_k(\thetavec_\Lambda)$, $k=1,\ldots,K$, are nonsingular matrices, and, thus, \eqref{eq:vec_covariance_inequality} is well defined. By substituting $\xsivec(\xvec,\thetavec)=\hat{\thetavec}_{\Lambda_k}\zp-\thetavec_{\Lambda_k}\zp-\bvec_k(\thetavec,{\Lambda})$ in \eqref{eq:vec_covariance_inequality},
we obtain
\beqna
\label{CS}
{\rm{E}}_{\thetavecsmall_{\Lambda}}\left[(
\hat{\thetavec}_{\Lambda_k}\zp-\thetavec_{\Lambda_k}\zp-\bvec_k(\thetavec,{\Lambda}))\right.\hspace{2cm}
\nonumber\\
\left. \times
(\hat{\thetavec}_{\Lambda_k}\zp-\thetavec_{\Lambda_k}\zp-\bvec_k(\thetavec,{\Lambda}))^T|{\hat{\Lambda}=\Lambda}_k\right]\nonumber\\
\succeq 
\Lambdamat_k(\thetavec_\Lambda)\Jmat_k^{-1}(\thetavec_{\Lambda})\Lambdamat_k^T(\thetavec_\Lambda),\hspace{1cm}
\eeqna
where
\beqna
\label{Lambda_def}
\Lambdamat_k(\thetavec_\Lambda)\hspace{6.75cm}\nonumber\\
\define {\rm{E}}_{\thetavecsmall_{\Lambda}}\left[\left(
\hat{\thetavec}_{\Lambda_k}\zp-\thetavec_{\Lambda_k}\zp-\bvec_k(\thetavec,{\Lambda})\right)\upsilonvec_k^T(\xvec,\thetavec_{\Lambda})|{\hat{\Lambda}=\Lambda}_k\right]
\nonumber\\
= {\rm{E}}_{\thetavecsmall_{\Lambda}}\left[\left(
\hat{\thetavec}_{\Lambda_k}\zp-\thetavec_{\Lambda_k}\zp\right)\upsilonvec_k^T(\xvec,\thetavec_{\Lambda})|{\hat{\Lambda}=\Lambda}_k\right],\hspace{1.6cm}
\eeqna
is an $M\times |\Lambda|$ matrix,
and  the last equality in \eqref{Lambda_def} is obtained by substituting \eqref{zero_val}.
It can be seen that
the $m$th element of the estimation error vector satisfies
\beqna
\left[
\hat{\thetavec}_{\Lambda_k}\zp-\thetavec_{\Lambda_k}\zp\right]_m=\left\{\begin{array}{lr}
\hat{\theta}_m-\theta_m,& m\in\Lambda_k\\
		0,& m\in\Lambda_k^c 
\end{array}\right. ,
\eeqna
 $\forall m=1,\ldots,M$.
Thus, 
\be
\label{zero_app}
\left[\Lambdamat_k(\thetavec_{\Lambda})\right]_{m,l}=0,~ m\in\Lambda_k^c .
\ee
Otherwise,
 by using (\ref{Lambda_def}) and
integration by parts, and assuming   Condition \ref{cond2}, it can be verified that
\beqna
\label{by_parts}
\left[\Lambdamat_k(\thetavec_{\Lambda})\right]_{m,l}=
\int_{{\mathcal{A}}_{k}}
(
\hat{\theta}_{m}-\theta_{m})
\frac{\partial f(\xvec|{\hat{\Lambda}=\Lambda}_k;\thetavec_{\Lambda})}{\partial [\thetavec_{\Lambda}]_l}{\ud}\xvec\hspace{0.6cm}
\nonumber\\=
\frac{\partial}{\partial [\thetavec_{\Lambda}]_l}\left\{{\rm{E}}_{\thetavecsmall_{\Lambda}}\left[\left.\hat{\theta}_m-\theta_m
\right|{\hat{\Lambda}=\Lambda}_k\right]
+\theta_m\right\}
\nonumber\\
=\frac{\partial}{\partial [\thetavec_{\Lambda}]_l}\left[\bvec_k(\thetavec,\Lambda)\right]_{m}
+\delta_{l,m}\hspace{2.25cm}
\nonumber\\
=[\Gmat_k(\thetavec,\Lambda)]_{m,\Lambda_l}+\delta_{l,m},\hspace{2.75cm}
\eeqna
$\forall l=1,\ldots,|\Lambda|$
 and $m\in\Lambda_k$, 
where the third equality is obtained 
by using the   selective bias  definition from (\ref{11_bias})
and the last equality is obtained by substituting \eqref{der_bias}.
Thus, \eqref{zero_app} and \eqref{by_parts} imply that
\be
\label{I}
\Lambdamat_k(\thetavec_{\Lambda})=
\Dmat_{k}(\Lambda)
+\Gmat_k(\thetavec,\Lambda),
\ee
where $\Dmat_{k}(\Lambda)$ is defined in (\ref{Imat}).
By substituting (\ref{I}) and \eqref{11_bias} in (\ref{CS}), we obtain
\beqna
\label{CS_k}
{\rm{E}}_{\thetavecsmall_{\Lambda}}\left[\left.(
\hat{\thetavec}_{\Lambda_k}\zp-\thetavec_{\Lambda_k}\zp)
(\hat{\thetavec}_{\Lambda_k}\zp-\thetavec_{\Lambda_k}\zp)^T\right|{\hat{\Lambda}=\Lambda}_k\right]\hspace{1cm}
\nonumber\\-	\bvec_k(\thetavec,\Lambda)	\bvec_k^T(\thetavec,\Lambda)
\succeq \Psi_k(\thetavec,\Lambda),
\eeqna
$k=1,\ldots,K$,
where $\Psi_k(\thetavec,\Lambda)$ is defined in \eqref{psi_def}.
By  multiplying  (\ref{CS_k}) by $\pi_{k}(\thetavec_{\Lambda})$ and  summing over the candidate models,  $k=1,\ldots,K$, we obtain 
 the selective CRB  in (\ref{CRB})-\eqref{BsCRB}.
Furthermore, by substituting (\ref{CRB})  in the relation in \eqref{MSE}
we obtain  the MSE lower bound in \eqref{bound3_new}.

%%%%%%%%%%%%%%%%555
	\section{Proof of Lemma \ref{lemma1}}
	\label{proofLemma}
Under Conditions \ref{cond1} and \ref{cond2}, 
\eqref{zero_val} is satisfied. 
Additionally, under Conditions \ref{cond3}  and \ref{cond4},
by using \eqref{l_define}, \eqref{JJJdef}, and the product rule twice on the  selective FIM from
(\ref{JJJdef}),
we obtain
\beqna
\label{thanks1}
\Jmat_k(\thetavec_\Lambda)= {\rm{E}}_{\thetavecsmall_\Lambda}\left[\left.
\nabla_{\thetavecsmall_{\Lambda}} \log f(\xvec|{\hat{\Lambda}=\Lambda}_k;\thetavec_{\Lambda})\hspace{1.75cm}\right.\right.
\nonumber\\\times
\left.\left.
\nabla_{\thetavecsmall_{\Lambda}}^T\log f(\xvec|{\hat{\Lambda}=\Lambda}_k;\thetavec_{\Lambda}) \right|{\hat{\Lambda}=\Lambda}_k
\right]\hspace{0.65cm}
 \nonumber\\
 =\nabla_{\thetavecsmall_{\Lambda}} {\rm{E}}_{\thetavecsmall_\Lambda}\left[\left.
\nabla_{\thetavecsmall_{\Lambda}}^T\log f(\xvec|{\hat{\Lambda}=\Lambda}_k;\thetavec_{\Lambda})\right|{\hat{\Lambda}=\Lambda}_k\right] \nonumber\\
 -{\rm{E}}_{\thetavecsmall_\Lambda}\left[\left.
\nabla_{\thetavecsmall_{\Lambda}}^2 \log f(\xvec|{\hat{\Lambda}=\Lambda}_k;\thetavec_{\Lambda}) \right|{\hat{\Lambda}=\Lambda}_k\right]
\nonumber\\=
 -{\rm{E}}_{\thetavecsmall_\Lambda}\left[\left.
\nabla_{\thetavecsmall_{\Lambda}}^2 \log f(\xvec|{\hat{\Lambda}=\Lambda}_k;\thetavec_{\Lambda}) \right|{\hat{\Lambda}=\Lambda}_k\right],
\eeqna
where the last equality is obtained by substituting  \eqref{zero_val}.
By substituting  \eqref{bayes} in (\ref{thanks1}), we obtain (\ref{q}).

Moreover,
 by substituting (\ref{l_define2}) in
(\ref{JJJdef}), one obtains
\beqna
\label{JJJdef_re}
\Jmat_k(\thetavec_\Lambda) \hspace{5.75cm}\nonumber\\= {\rm{E}}_{\thetavecsmall_\Lambda}\left[\left.
 \nabla_{\thetavecsmall_{\Lambda}} \log f(\xvec;\thetavec_\Lambda)
\nabla_{\thetavecsmall_{\Lambda}}^T\log f(\xvec;\thetavec_\Lambda) \right|{\hat{\Lambda}=\Lambda}_k
\right]
\nonumber\\- \nabla_{\thetavecsmall_{\Lambda}} \log\pi_{k}(\thetavec_\Lambda)
{\rm{E}}_{\thetavecsmall_\Lambda}\left[\left.\nabla_{\thetavecsmall_{\Lambda}}^T \log f(\xvec;\thetavec_\Lambda) \right|{\hat{\Lambda}=\Lambda}_k
\right]
\nonumber\\-{\rm{E}}_{\thetavecsmall_\Lambda}\left[\left.
 \nabla_{\thetavecsmall_{\Lambda}} \log f(\xvec;\thetavec_{\Lambda})\right|{\hat{\Lambda}=\Lambda}_k
\right]\nabla_{\thetavecsmall_{\Lambda}}^T \log\pi_{k}(\thetavec_\Lambda)
\nonumber\\+
 \nabla_{\thetavecsmall_{\Lambda}} \log \pi_{k}(\thetavec_\Lambda)
\nabla_{\thetavecsmall_{\Lambda}}^T\log\pi_{k}(\thetavec_\Lambda).\hspace{2.4cm}
\eeqna
Since ${\mathcal{A}}_{k}$ is independent of $\thetavec$,
and by using regularity condition \ref{cond2},
it can be noticed that
\beqna
\label{nine}
\nabla_{\thetavecsmall_\Lambda} \log\pi_{k}(\thetavec_\Lambda)=
\nabla_{\thetavecsmall_\Lambda} \log \Pr({\hat{\Lambda}=\Lambda}_k;\thetavec_\Lambda)\hspace{1.35cm}
\nonumber\\=
\frac{\nabla_{\thetavecsmall_\Lambda} \Pr({\hat{\Lambda}=\Lambda}_k;\thetavec_\Lambda)}{\Pr({\hat{\Lambda}=\Lambda}_k;\thetavec_\Lambda)}\hspace{1.8cm}
\nonumber\\
={\rm{E}}_{\thetavecsmall_\Lambda}\left[\left.
\nabla_{\thetavecsmall_\Lambda} \log f(\xvec;\thetavec_\Lambda) \right|{\hat{\Lambda}=\Lambda}_k
\right].
\eeqna
Substitution of 
\eqref{nine} in \eqref{JJJdef_re} results in  \eqref{emp_Jk}.
%%%%
Similarly to the contents of this Appendix, derivations of alternative formulations of the conditional FIM can be found in \cite{Chaumette2005,RoutTong2016}.
%%%%%%%%%%%%%%%%

%%%%%%%%%%%%%%%%%
	\section{Proof of Theorem \ref{Th2}}
	\label{sparseAPP}
	We consider the model from (\ref{sparseMeas}), where $\wvec$ is a zero-mean Gaussian vector with a covariance matrix $\sigma^2\Imat_L$
	and with the OST selection rule.
It can be seen that for the OST selection rule from (\ref{sparseSelection}) and under the Gaussian distribution,
the probability of the $m$th index exceeding the  level $c$
is
\beqna
\label{foldedCDF}
	\Pr\left(|\avec_m^T\xvec|>c;\thetavec_{\Lambda}\right)
	=1-\Phi(\alpha_m)+\Phi(\beta_m),
\eeqna
$\forall m=1,\ldots,M$.
Thus, it follows that under the i.i.d. Gaussian  noise assumption
the probability of selecting the $k$th model  with the support set $\Lambda_k$ by the OST rule is given by
\beqna
\label{sparseProof5}
\pi_k(\thetavec_{\Lambda})\hspace{6.5cm}
\nonumber\\=
\prod_{l\in\Lambda_k}\Pr\left(|\avec_l^T\xvec|\geq c;\thetavec_{\Lambda}\right)
\prod_{m\notin\Lambda_k}\Pr\left(|\avec_m^T\xvec|<c;\thetavec_{\Lambda}\right)\nonumber\\=
\prod_{l\in\Lambda_k}
\left(1-\Phi(\alpha_l)+\Phi(\beta_l)\right)
\prod_{m\notin\Lambda_k}\Phi(\alpha_m)-\Phi(\beta_m),\hspace{0.4cm}
\eeqna
for any  $k=1,\ldots,K$,
where the second equality is obtained by substituting \eqref{foldedCDF}.
In addition, the  model from (\ref{sparseMeas}) 
implies that 
\be\label{pdf_g}
f(\xvec;\thetavec_\Lambda)=
\frac{1}{(2\pi \sigma^2)^{\frac{L}{2}}} e^{-\frac{1}{2\sigma^2}||\xvec -\Amat_\Lambda\thetavecsmall_\Lambda||^2}.
\ee
By substituting \eqref{sparseProof5} and \eqref{pdf_g} in \eqref{bayes}, it can be verified that in this case
 $	f(\xvec |{\hat{\Lambda}=\Lambda}_k; \thetavec_\Lambda)$ is a smooth function in the sense that its first- and second-order derivatives are well defined. In addition, the support of the conditional pdf, $f(\xvec|{\hat{\Lambda}=\Lambda}_k;\thetavec_{\Lambda}) $, is $\mathbb{R}^{L}$
 for all   $\thetavec_\Lambda$, which implies that Condition \ref{cond2} holds.
Thus, 
	 Conditions \ref{cond1}-\ref{cond4} are satisfied  for the considered model as long as the selective FIMs,
		$\Jmat_k(\thetavec_\Lambda)$, $k=1,\ldots,K$, are  nonsingular matrices, as required in Condition \ref{cond1}.

Under Conditions  \ref{cond3} and \ref{cond4},
and by using \eqref{pdf_g},
it can be verified that 
\be
\label{aaa}
\nabla_{\thetavecsmall_{\Lambda}}(\nabla^T_{\thetavecsmall_{\Lambda}}\log f(\xvec;\thetavec_\Lambda))
=-\frac{\Amat_\Lambda^T\Amat_\Lambda}{\sigma^2}.
\ee
Since the matrix on the r.h.s. of \eqref{aaa} is a deterministic matrix, we can conclude that
\be
\label{sparseCRB}
	{\rm{E}}_{\thetavecsmall_\Lambda}[\nabla_{\thetavecsmall_{\Lambda}}(\nabla^T_{\thetavecsmall_{\Lambda}}\log f(\xvec;\thetavec_\Lambda))|{\hat{\Lambda}=\Lambda}_k]=-\frac{\Amat_\Lambda^T\Amat_\Lambda}{\sigma^2},
\ee
$\forall k=1,\ldots,K$.
In addition, 
the derivative of the log of the probability $\pi_k(\thetavec_{\Lambda})$ from
\eqref{sparseProof5} w.r.t. $\theta_s$ 
  is 
\beqna
\label{SparseDer1}
\frac{\partial}{\partial\theta_s}\log\pi_{k}(\thetavec_{\Lambda})
=
\sum_{l\in\Lambda_k}\frac{\avec_l^T\avec_s}{\sigma}\frac{\phi(\alpha_l)
-\phi(\beta_l)}
{1-\Phi(\alpha_l)+\Phi(\beta_l)}
\nonumber\\
+\sum_{m\notin\Lambda_k}\frac{\avec_m^T\avec_s}{\sigma}
\frac{-\phi(\alpha_m)
+\phi(\beta_m)}{\Phi(\alpha_m)-\Phi(\beta_m)}
\eeqna
for any $s\in\Lambda$, where we use 
\beqna
\label{Phi_der}
\frac{\partial}{\partial\theta_s}
\Phi(\alpha_m)
=-\frac{\avec_m^T\avec_s}{\sigma}\phi(\alpha_m), \forall m=1,\ldots,M.
\eeqna
Similarly, by using \eqref{SparseDer1}, \eqref{Phi_der}, and 
\beqna
\frac{\partial}{\partial\theta_t}
\phi(\alpha_m)
=\frac{\avec_m^T\avec_t}{\sigma}\phi(\alpha_m)\alpha_m,
\eeqna
 it can be verified that
the second-order derivatives of \eqref{sparseProof5} w.r.t. the general parameters $\theta_s$ and $\theta_t$,
  $t,s\in\Lambda$, are given by 
\beqna
\label{SparseDer2}
\frac{\partial^2\log \pi_k(\thetavec_{\Lambda})}{\partial \theta_s \partial \theta_t}
=\avec_s^T \Qmat_k \avec_t,
\eeqna
where the matrix $\Qmat_k$ is defined in \eqref{Q_def}.
Thus,
\beqna
\label{SparseDer8}
\nabla_{\thetavecsmall_{\Lambda}}\nabla^T_{\thetavecsmall_{\Lambda}}{\rm{E}}_{\thetavecsmall_\Lambda} \log \pi_k(\thetavec_{\Lambda})
=\Amat_{\Lambda}^T \Qmat_k \Amat_{\Lambda},
\eeqna
		for any $s=1,\ldots,|\Lambda|$.
Now, by substituting \eqref{sparseCRB} and \eqref{SparseDer8} in (\ref{q}),
which is equivalent to \eqref{JJJdef},
 we obtain  the selective FIM for the Gaussian case with OST threshold in \eqref{spsJ}.

\bibliographystyle{IEEEbib}

\begin{thebibliography}{10}

\bibitem{Sando_Mitra_Stoica2002}
S.~Sando, A.~Mitra, and P.~Stoica,
\newblock ``On the {C}ram$\acute{\text{e}}$r-{R}ao bound for model-based
  spectral analysis,''
\newblock {\em IEEE Signal Process. Lett.}, vol. 9, no. 2, pp. 68--71, Feb.
  2002.

\bibitem{multivariate}
S.~L. Sclove,
\newblock ``Application of model-selection criteria to some problems in
  multivariate analysis,''
\newblock {\em Psychometrika}, vol. 52, no. 3, pp. 333--343, Sep. 1987.

\bibitem{7006732}
Y.~Doweck, A.~Amar, and I.~Cohen,
\newblock ``Joint model order selection and parameter estimation of chirps with
  harmonic components,''
\newblock {\em IEEE Trans. Signal Process.}, vol. 63, no. 7, pp. 1765--1778,
  Apr. 2015.

\bibitem{stoica2004model}
P.~Stoica and Y.~Selen,
\newblock ``Model-order selection: a review of information criterion rules,''
\newblock {\em IEEE Signal Processing Magazine}, vol. 21, no. 4, pp. 36--47,
  2004.

\bibitem{BnA}
K.~P. Burnham and D.R. Anderson,
\newblock {\em Model Selection and Multimodel Inference: A Practical
  Information-Theoretic Approach},
\newblock Springer New York, 2003.

\bibitem{WaxKailath}
M.~Wax and T.~Kailath,
\newblock ``Detection of signals by information theoretic criteria,''
\newblock {\em IEEE Trans. Acoustics, Speech, and Signal Process.}, vol. 33,
  no. 2, pp. 387--392, Apr. 1985.

\bibitem{Nadler}
B.~Nadler and A.~Kontorovich,
\newblock ``Model selection for sinusoids in noise: Statistical analysis and a
  new penalty term,''
\newblock {\em IEEE Trans. Signal Process.}, vol. 59, no. 4, pp. 1333--1345,
  Apr. 2011.

\bibitem{Ottersten1993}
B.~Ottersten, M.~Viberg, P.~Stoica, and A.~Nehorai,
\newblock {\em Exact and Large Sample Maximum Likelihood Techniques for
  Parameter Estimation and Detection in Array Processing}, pp. 99--151,
\newblock Springer, Berlin, Heidelberg, 1993.

\bibitem{Bethel_Bell_2004}
R.~E. Bethel and K.~L. Bell,
\newblock ``Maximum likelihood approach to joint array detection/estimation,''
\newblock {\em IEEE Trans. Aerospace and Electronic Systems}, vol. 40, no. 3,
  pp. 1060--1072, July 2004.

\bibitem{Nikolic_Nehorai_Djordjevic2012}
M.~M. Nikolic, A.~Nehorai, and A.~R. Djordjevic,
\newblock ``Estimation of direction of arrival using multipath on array
  platforms,''
\newblock {\em IEEE Trans. Antennas and Propagation}, vol. 60, no. 7, pp.
  3444--3454, July 2012.

\bibitem{AIC}
H.~Akaike,
\newblock ``{A new look at the statistical model identification},''
\newblock {\em IEEE Trans. Automatic Control}, vol. 19, no. 6, pp. 716--723,
  Dec. 1974.

\bibitem{MDL}
J.~Rissanen,
\newblock ``A universal prior for integers and estimation by minimum
  description length,''
\newblock {\em The Annals of Statistics}, vol. 11, no. 2, pp. 416--431, 1983.

\bibitem{hurvich1989regression}
C.~M. Hurvich and C.~L. Tsai,
\newblock ``Regression and time series model selection in small samples,''
\newblock {\em Biometrika}, vol. 76, no. 2, pp. 297--307, 1989.

\bibitem{Francos1}
M.~Kliger and J.~M. Francos,
\newblock ``Strong consistency of the over-and underdetermined {LSE} of 2{D}
  exponentials in white noise,''
\newblock {\em IEEE Trans. Inf. Theory}, vol. 51, no. 9, pp. 3314--3321, 2005.

\bibitem{tropp2006}
J.~A. Tropp,
\newblock ``{Just relax: Convex programming methods for identifying sparse
  signals in noise},''
\newblock {\em IEEE Trans. Inf. Theory}, vol. 52, no. 3, pp. 1030--1051, 2006.

\bibitem{candes2007}
E.~Candes and T.~Tao,
\newblock ``The {D}antzig selector: Statistical estimation when p is much
  larger than n,''
\newblock {\em The Annals of Statistics}, vol. 35, no. 6, pp. 2313--2351, 2007.

\bibitem{donoho2003optimally}
D.~L. Donoho and M.~Elad,
\newblock ``Optimally sparse representation in general (nonorthogonal)
  dictionaries via $\ell$-1 minimization,''
\newblock {\em Proceedings of the National Academy of Sciences}, vol. 100, no.
  5, pp. 2197--2202, 2003.

\bibitem{CS_Davenport_Eldar}
M.~A. Davenport, M.~F. Duarte, Y.~C. Eldar, and G.~Kutyniok,
\newblock {\em Introduction to Compressed Sensing},
\newblock Cambridge University Press, 2012,
\newblock Compressed Sensing: Theory and Applications, Edited by Y. C. Eldar
  and G. Kutyniok, Cambridge University Press.

\bibitem{OMP_Mallat93}
S.~Mallat and Z.~Zhang,
\newblock ``Matching pursuit with time-frequency dictionaries,''
\newblock {\em IEEE Trans. Signal Process.}, vol. 41, pp. 3397--3415, Dec.
  1993.

\bibitem{Ye_Bresler_Moulin2003}
Y.~Bresler J.~C.~Ye and P.~Moulin,
\newblock ``{C}ram$\acute{\text{e}}$r-{R}ao bounds for parametric shape
  estimation in inverse problems,''
\newblock {\em IEEE Trans. Image Process.}, vol. 12, no. 1, pp. 71--84, Jan.
  2003.

\bibitem{sparse_con}
Z.~Ben-Haim and Y.~C. Eldar,
\newblock ``The {C}ram$\acute{\text{e}}$r-{R}ao bound for estimating a sparse
  parameter vector,''
\newblock {\em IEEE Trans. Signal Process.}, vol. 58, no. 6, pp. 3384--3389,
  June 2010.

\bibitem{ZhaoZhang}
R.~Berk, L.~Brown, A.~Buja, K.~Zhang, and L.~Zhao,
\newblock ``Valid post-selection inference,''
\newblock {\em The Annals of Statistics}, vol. 41, no. 2, pp. 802--837, Apr.
  2013.

\bibitem{potscher1991effects}
B.~M P$\ddot{\text{o}}$tscher,
\newblock ``Effects of model selection on inference,''
\newblock {\em Econometric Theory}, vol. 7, no. 2, pp. 163--185, 1991.

\bibitem{Leeb_Potscher2005}
H.~Leeb and B.~M. P$\ddot{\text{o}}$tscher,
\newblock ``Model selection and inference: Facts and fiction,''
\newblock {\em Econometric Theory}, vol. 21, no. 1, pp. 21--59, 2005.

\bibitem{athley2005threshold}
F.~Athley,
\newblock ``Threshold region performance of maximum likelihood direction of
  arrival estimators,''
\newblock {\em IEEE Trans. Signal Process.}, vol. 53, no. 4, pp. 1359--1373,
  2005.

\bibitem{Lehmann}
E.~L. Lehmann and J.~P. Romano,
\newblock {\em Testing {S}tatistical {H}ypotheses},
\newblock Springer Texts in Statistics, New York, 3nd edition, 2005.

\bibitem{Lee_Taylor_2014}
J.~D. Lee and J.~E. Taylor,
\newblock ``Exact post model selection inference for marginal screening,''
\newblock in {\em Advances in Neural Information Processing Systems}, 2014, pp.
  136--144.

\bibitem{EfronX}
B.~Efron,
\newblock ``Estimation and accuracy after model selection,''
\newblock {\em J. Am. Statist. Assoc.}, vol. 109, no. 507, pp. 991--1007, 2014.

\bibitem{Benjamini2005false}
Y.~Benjamini and D.~Yekutieli,
\newblock ``False discovery rate--adjusted multiple confidence intervals for
  selected parameters,''
\newblock {\em J. Am. Statist. Assoc.}, vol. 100, no. 469, pp. 71--81, 2005.

\bibitem{Kabaila_Leeb_2006}
P.~Kabaila and H.~Leeb,
\newblock ``On the large-sample minimal coverage probability of confidence
  intervals after model selection,''
\newblock {\em J. Am. Statist. Assoc.}, vol. 101, no. 474, pp. 619--629, 2006.

\bibitem{TibsAs}
R.~J. Tibshirani, A.~Rinaldo, R.~Tibshirani, and L.~Wasserman,
\newblock ``Uniform asymptotic inference and the bootstrap after model
  selection,''
\newblock {\em The Annals of Statistics}, vol. 46, no. 3, pp. 1255--1287, 2018.

\bibitem{Rosenblatt2014401}
J.D. Rosenblatt and Y.~Benjamini,
\newblock ``Selective correlations; not voodoo,''
\newblock {\em NeuroImage}, vol. 103, pp. 401--410, 2014.

\bibitem{Tibshirani_Taylor_Lockhart_Tibshirani2016}
R.~J. Tibshirani, J.~Taylor, R.~Lockhart, and R.~Tibshirani,
\newblock ``Exact post-selection inference for sequential regression
  procedures,''
\newblock {\em J. Am. Statist. Assoc.}, vol. 111, no. 514, pp. 600--620, 2016.

\bibitem{2014arXiv1410.2597F}
W.~Fithian, D.~Sun, and J.~Taylor,
\newblock ``Optimal inference after model selection,''
\newblock {\em ArXiv e-prints, \url{https://arxiv.org/abs/1410.2597}}, Oct.
  2014.

\bibitem{Heller_Meir_Chatterjee2017}
R.~{Heller}, A.~{Meir}, and N.~{Chatterjee},
\newblock ``{Post-selection estimation and testing following aggregated
  association tests},''
\newblock {\em ArXiv e-prints, \url{https://arxiv.org/abs/1711.00497}}, Nov.
  2017.

\bibitem{meir2017tractable}
A.~Meir and M.~Drton,
\newblock ``Tractable post-selection maximum likelihood inference for the
  {L}asso,''
\newblock {\em ArXiv e-prints, \url{https://arxiv.org/abs/1705.09417}}, p.
  arXiv:1705.09417, May 2017.

\bibitem{hero2}
E.~Bashan, R.~Raich, and A.~O. Hero,
\newblock ``Optimal two-stage search for sparse targets using convex
  criteria,''
\newblock {\em IEEE Trans. Signal Process.}, vol. 56, no. 11, pp. 5389--5402,
  Nov. 2008.

\bibitem{Harel_Routtenberg_Bayesian}
N.~Harel and T.~Routtenberg,
\newblock ``Bayesian post-model-selection estimation,''
\newblock {\em IEEE Signal Process. Lett.}, vol. 28, pp. 175--179, 2021.

\bibitem{Chaumette2005}
E.~Chaumette, P.~Larzabal, and P.~Forster,
\newblock ``On the influence of a detection step on lower bounds for
  deterministic parameter estimation,''
\newblock {\em IEEE Trans. Signal Process.}, vol. 53, no. 11, pp. 4080--4090,
  Nov. 2005.

\bibitem{energy}
E.~Chaumette and P.~Larzabal,
\newblock ``Cram$\acute{\text{e}}$r-{R}ao bound conditioned by the energy
  detector,''
\newblock {\em IEEE Signal Process. Lett.}, vol. 14, no. 7, pp. 477--480, July
  2007.

\bibitem{Weiss_Routtenberg_Messer}
T.~Weiss, T.~Routtenberg, and H.~Messer,
\newblock ``Total performance evaluation of intensity estimation after
  detection,''
\newblock {\em Accepted to: Signal Processing,
  \url{http://www.ee.bgu.ac.il/~tirzar/publications2.html}}, 2021.

\bibitem{Pakrooh_Scharf_Howard_2015}
P.~Pakrooh, A.~Pezeshki, L.~L. Scharf, D.~Cochran, and S.~D. Howard,
\newblock ``Analysis of {F}isher information and the
  {C}ram$\acute{\text{e}}$r-{R}ao bound for nonlinear parameter estimation
  after random compression,''
\newblock {\em IEEE Trans. Signal Process.}, vol. 63, no. 23, pp. 6423--6428,
  Dec. 2015.

\bibitem{Richmond2006}
C.~D. {Richmond},
\newblock ``Mean-squared error and threshold {SNR} prediction of
  maximum-likelihood signal parameter estimation with estimated colored noise
  covariances,''
\newblock {\em IEEE Trans. Inf. Theory}, vol. 52, no. 5, pp. 2146--2164, 2006.

\bibitem{richmond2012aspects}
C.~D. Richmond and L.~L. Horowitz,
\newblock ``Aspects of threshold region mean squared error prediction: Method
  of interval errors, bounds, {T}aylor's theorem and extensions,''
\newblock in {\em Asilomar Conference on Signals, Systems and Computers
  (ASILOMAR)}, 2012, pp. 13--17.

\bibitem{Merhav_2011}
N.~{Merhav},
\newblock ``Threshold effects in parameter estimation as phase transitions in
  statistical mechanics,''
\newblock {\em IEEE Transactions on Information Theory}, vol. 57, no. 10, pp.
  7000--7010, 2011.

\bibitem{Routtenberg_Tong_ICASSP14}
T.~Routtenberg and L.~Tong,
\newblock ``The {C}ram$\acute{\text{e}}$r-{R}ao bound for
  estimation-after-selection,''
\newblock in {\em IEEE International Conference on Acoustics, Speech, and
  Signal Processing (ICASSP)}, May 2014, pp. 414--418.

\bibitem{RoutTong2016}
T.~Routtenberg and L.~Tong,
\newblock ``Estimation after parameter selection: Performance analysis and
  estimation methods,''
\newblock {\em IEEE Trans. Signal Process.}, vol. 64, no. 20, pp. 5268--5281,
  Oct. 2016.

\bibitem{Harel_Routtenberg_2019}
N.~Harel and T.~Routtenberg,
\newblock ``Low-complexity methods for estimation after parameter selection,''
\newblock {\em IEEE Trans. Signal Process.}, vol. 68, pp. 1152--1167, 2020.

\bibitem{richmondHorowitz}
C.~D. Richmond and L.~L. Horowitz,
\newblock ``Parameter bounds on estimation accuracy under model
  misspecification,''
\newblock {\em IEEE Trans. Signal Process.}, vol. 63, no. 9, pp. 2263--2278,
  May 2015.

\bibitem{vuong}
Q.~H. Vuong,
\newblock ``Cram$\acute{\text{e}}$r-{R}ao bounds for misspecified models,''
\newblock {\em {working paper 652, Div. of the Humanities and Social Sci.,
  Caltech, Pasadena, USA}}, Feb. 1986.

\bibitem{pajovic}
M.~Pajovic,
\newblock ``Misspecified {B}ayesian {C}ram$\acute{\text{e}}$r-{R}ao bound for
  sparse {B}ayesian learning,''
\newblock in {\em Statistical Signal Processing Workshop (SSP)}, June 2018, pp.
  263--267.

\bibitem{Fortunati_Gini_Greco2016}
S.~Fortunati, F.~Gini, and M.~S. Greco,
\newblock ``The misspecified {C}ram$\acute{\text{e}}$r-{R}ao bound and its
  application to scatter matrix estimation in complex elliptically symmetric
  distributions,''
\newblock {\em IEEE Trans. Signal Process.}, vol. 64, no. 9, pp. 2387--2399,
  May 2016.

\bibitem{gorman1990}
J.~D. Gorman and A.~O. Hero,
\newblock ``Lower bounds for parametric estimation with constraints,''
\newblock {\em IEEE Trans. Inf. Theory}, vol. 36, no. 6, pp. 1285--1301, 1990.

\bibitem{stoica1998}
P.~Stoica and B.~C. Ng,
\newblock ``{On the Cram$\acute{\text{e}}$r-{R}ao bound under parametric
  constraints},''
\newblock {\em IEEE Signal Process. Lett.}, vol. 5, no. 7, pp. 177--179, 1998.

\bibitem{babadi2009}
B.~Babadi, N.~Kalouptsidis, and V.~Tarokh,
\newblock ``Asymptotic achievability of the {C}ram$\acute{\text{e}}$r-{R}ao
  bound for noisy compressive sampling,''
\newblock {\em IEEE Trans. Signal Process.}, vol. 57, no. 3, pp. 1233--1236,
  2009.

\bibitem{meirRoutSSP}
E.~Meir and T.~Routtenberg,
\newblock ``Selective {C}ram$\acute{\text{e}}$r-{R}ao bound for estimation
  after model selection,''
\newblock in {\em Statistical Signal Processing Workshop (SSP)}, June 2018, pp.
  757--761.

\bibitem{Kay_estimation}
S.~M. Kay,
\newblock {\em Fundamentals of Statistical Signal Processing: Estimation
  Theory},
\newblock Prentice Hall, 1993.

\bibitem{RoutPhd}
T.~Routtenberg,
\newblock {\em ``Parameter Estimation under Arbitrary Cost Functions with
  Constraints"},
\newblock Ph.D. thesis,
  \url{http://aranne5.bgu.ac.il/others/RouttenbergTirza3.pdf}, Ben-Gurion
  University of the Negev, May 2012.

\bibitem{PCRB_J}
T.~Routtenberg and J.~Tabrikian,
\newblock ``Non-{B}ayesian periodic {C}ram$\acute{\text{e}}$r-{R}ao bound,''
\newblock {\em IEEE Trans. Signal Process.}, vol. 61, no. 4, pp. 1019--1032,
  Feb. 2013.

\bibitem{Routtenberg_cyclic}
T.~Routtenberg and J.~Tabrikian,
\newblock ``Cyclic {B}arankin-type bounds for non-{B}ayesian periodic parameter
  estimation,''
\newblock {\em IEEE Trans. Signal Process.}, vol. 62, no. 13, pp. 3321--3336,
  July 2014.

\bibitem{Eyal_constraint}
E.~Nitzan, T.~Routtenberg, and J.~Tabrikian,
\newblock ``Cram$\acute{\text{e}}$r-{R}ao bound for constrained parameter
  estimation using {L}ehmann-unbiasedness,''
\newblock {\em IEEE Trans. Signal Process.}, vol. 67, no. 3, pp. 753--768, Feb.
  2019.

\bibitem{SomekhBaruch_Leshem2017}
A.~Somekh-Baruch, A.~Leshem, and V.~Saligrama,
\newblock ``On the non-existence of unbiased estimators in constrained
  estimation problems,''
\newblock {\em IEEE Trans. Inf. Theory}, vol. PP, no. 99, pp. 1--1, 2017.

\bibitem{Theory_Statistics_book}
Mark~J. Schervish,
\newblock {\em Theory of Statistics},
\newblock Springer-Verlag New York, 1995.

\bibitem{white1982maximum}
H.~White,
\newblock ``Maximum likelihood estimation of misspecified models,''
\newblock {\em Econometrica: Journal of the Econometric Society}, pp. 1--25,
  1982.

\bibitem{BARANKIN}
E.~W. Barankin,
\newblock ``Locally best unbiased estimates,''
\newblock {\em Ann. Math. Stat.}, vol. 20, pp. 477--501, 1946.

\bibitem{berisha2015empirical}
V.~Berisha and A.~O. Hero,
\newblock ``Empirical non-parametric estimation of the {F}isher information,''
\newblock {\em IEEE Signal Process. Lett.}, vol. 22, no. 7, pp. 988--992, 2015.

\bibitem{spall2005monte}
J.~C. Spall,
\newblock ``Monte {C}arlo computation of the {F}isher information matrix in
  nonstandard settings,''
\newblock {\em Journal of Computational and Graphical Statistics}, vol. 14, no.
  4, pp. 889--909, 2005.

\bibitem{robert2013monte}
C.~Robert and G.~Casella,
\newblock {\em Monte Carlo statistical methods},
\newblock Springer Science \& Business Media, 2013.

\bibitem{donoho1994ideal}
D.~L. Donoho and J.~M. Johnstone,
\newblock ``Ideal spatial adaptation by wavelet shrinkage,''
\newblock {\em biometrika}, vol. 81, no. 3, pp. 425--455, 1994.

\bibitem{slavakis2013generalized}
Slavakis K., Kopsinis Y., Theodoridis S., and McLaughlin S.,
\newblock ``Generalized thresholding and online sparsity-aware learning in a
  union of subspaces,''
\newblock {\em IEEE Trans. Signal Process.}, vol. 61, no. 15, pp. 3760--3773,
  2013.

\bibitem{WassermanOST}
C.~R. Genovese, J.~Jin, L.~Wasserman, and Z.~Yao,
\newblock ``A comparison of the lasso and marginal regression,''
\newblock {\em Journal of Machine Learning Research}, vol. 13, no. Jun, pp.
  2107--2143, 2012.

\bibitem{OST}
W.~U. Bajwa and A.~Pezeshki,
\newblock ``Finite frames for sparse signal processing,''
\newblock in {\em Finite Frames}, pp. 303--335. Springer, 2013.

\bibitem{graybill1976theory}
F.~A. Graybill,
\newblock {\em Theory and application of the linear model},
\newblock Number 04; QA279, G7. 1976.

\bibitem{djuric1998asymptotic}
P.~M. Djuric,
\newblock ``Asymptotic {MAP} criteria for model selection,''
\newblock {\em IEEE Trans. Signal Process.}, vol. 46, no. 10, pp. 2726--2735,
  1998.

\bibitem{Wiesel_Eldar_Yeredor2008}
A.~Wiesel, Y.~C. Eldar, and A.~Yeredor,
\newblock ``Linear regression with {G}aussian model uncertainty: Algorithms and
  bounds,''
\newblock {\em IEEE Trans. Signal Process.}, vol. 56, no. 6, pp. 2194--2205,
  June 2008.

\bibitem{5701798}
Q.~Ding and S.~Kay,
\newblock ``Inconsistency of the {MDL}: On the performance of model order
  selection criteria with increasing signal-to-noise ratio,''
\newblock {\em IEEE Trans. Signal Process.}, vol. 59, no. 5, pp. 1959--1969,
  May 2011.

\bibitem{brychkov2012some}
Y.~A. Brychkov,
\newblock ``On some properties of the marcum {Q} function,''
\newblock {\em Integral Transforms and Special Functions}, vol. 23, no. 3, pp.
  177--182, 2012.

\bibitem{johnson1995continuous}
N.L. Johnson, S.~Kotz, and N.~Balakrishnan,
\newblock {\em Continuous univariate distributions},
\newblock Number v. 2 in Wiley series in probability and mathematical
  statistics: Applied probability and statistics. Wiley \& Sons, 1995.

\bibitem{VAN-TREES}
H.~L.~Van Trees and K.~L. Bell,
\newblock {\em Bayesian Bounds for Parameter Estimation and Nonlinear
  Filtering/Tracking},
\newblock Wiley-IEEE Press, 2007.

\bibitem{point_est}
E.~L. Lehmann and G.~Casella,
\newblock {\em Theory of Point Estimation (Springer Texts in Statistics)},
\newblock Springer, 2nd edition, 1998.

\end{thebibliography}

\end{document}